\documentclass[useAMS,usegraphicx,usenatbib,a4paper]{mn2e}
\usepackage[T1]{fontenc}
\usepackage{aecompl}
\usepackage[]{amsmath}
\usepackage{mathrsfs}
\usepackage{aas_macros}
\usepackage{color}
\usepackage[dvips, bookmarks, pdfborder={0 0 0.1}]{hyperref}
\newcommand{\VR}{\vec{r}}
\newcommand{\VV}{\vec{v}}
\newcommand{\vr}{v_{\rm r}}
\newcommand{\vt}{v_{\rm t}}
\newcommand{\D}{\mathrm{d}}

\newcommand{\msun}{\mathrm{M}_\odot}
\graphicspath{{figures/}}

\newcommand{\myplot}[1]{\includegraphics[width=0.5\textwidth]{#1}}
\newcommand{\myplottwo}[2]{\myplot{#1}\myplot{#2}}

\newcommand{\mytab}{\begin{table}[htb]}
\newcommand{\myfig}{\begin{figure}[htbp]}

\newcommand{\mybibstyle}{mymn}

\newcommand{\hl}[1]{#1}  
\newcommand{\cl}[1]{#1}

\begin{document}
\title[Dynamical state of Aquarius haloes]{The orbital PDF: the
dynamical state of Milky Way sized haloes and the intrinsic
uncertainty in the determination of their masses} \author[J. Han et
al.]  {Jiaxin Han,$^{1}$\thanks{jiaxin.han@durham.ac.uk}, Wenting
Wang,$^1$, Shaun Cole,$^1$, Carlos S. Frenk,$^1$\\ $^1$Institute for
Computational Cosmology, Department of Physics, University of Durham,
South Road, Durham, DH1 3LE\\ } \maketitle

\begin{abstract} 
Using realistic cosmological simulations of Milky Way sized haloes, we study their dynamical state and the accuracy of inferring their
mass profiles with steady-state models of dynamical tracers.
We use a new method that describes the phase-space distribution of a steady-state tracer population
in a spherical potential without any assumption regarding the distribution of their orbits. 
 Applying the method to five haloes from the Aquarius $\Lambda$CDM N-body simulation, 
 we find that dark matter particles are an accurate
 tracer that enables the halo mass and concentration parameters to
 be recovered with an accuracy of $5\%$.
 Assuming a potential profile of the NFW form
 does not significantly affect the fits in most cases, except for
 halo A whose density profile differs significantly from the NFW form, leading
 to a $30\%$ bias in the dynamically fitted parameters. The existence
 of substructures in the dark matter tracers only affects the fits by
 $\sim 1\%$. Applying the method to mock stellar haloes generated by a
 particle-tagging technique, we find the stars are farther from equilibrium
 than dark matter particles, yielding a systematic bias of $\sim 20\%$
 in the inferred mass and concentration parameter. The level of
 systematic biases obtained from a conventional distribution function
 fit to stars is comparable to ours, while similar fits to DM tracers
 are significantly biased in contrast to our fits. In line with
 previous studies, the mass bias is much reduced near the tracer half-mass radius.
\end{abstract}


\section{Introduction}
Dynamical modelling is of fundamental importance in the determination
of the mass distribution of dark matter haloes. To constrain the total
mass distribution or the gravitational potential, a large family of
dynamical methods work by fitting a proposed potential-dependent
distribution function (DF) to the observed phase-space distribution of a
tracer population. In such modelling, one should make as few assumptions as
possible so as to avoid biasing the results. In practice, a required minimal
assumption is that the system is in steady state, so that modelling
the tracer DF with a single observational snapshot is informative
without requiring the observation to take place at any special
moment. However, most existing methods involve additional assumptions,
for example, about the distribution of orbits, the functional form of
the DF, or the spatial distribution of tracer particles outside the
observational window. In a previous paper~\citep[][hereafter
Paper~I]{paperI}, we developed a method that can be used to
infer the potential while only making the assumption
that the tracer population is in a steady state.
In particular, taking a spherical potential as an example, we have shown
that the steady-state property translates into a fundamental orbital
Probability Density Function (oPDF), which provides enough information
to enable the inference of the halo potential. Applying this method to a set
of steady-state tracers in an NFW potential generated from Monte-Carlo
simulations, we showed that the method is able to recover the true
potential. While spherical symmetry is assumed, all the steps of the
method can be generalized to non-spherical cases.

A realistic halo from cosmological simulations or one in the real
universe may violate the assumptions of our method in
several ways. For example, spherical symmetry is only approximate
since we know haloes are triaxial~\citep{Frenk88,Jing02}. Also, the
potential and the distribution of any tracer
are not strictly static as haloes evolve with
time. Finally, real haloes are not smooth structures, since they
contain many subhaloes. In this work, we apply the oPDF method to
the dynamical distribution of dark matter and star particles in
simulated haloes, to explore the extent to which the tracers in a real
halo satisfy our model assumptions.

One important motivation for this work is to provide a generic
assessment of what to expect for the accuracy of dynamical mass
estimates of the Milky Way (MW) halo. The mass of the MW plays a
crucial role in interpreting many of the Local Group
observations~\citep{Jie,Kennedy,Cautun}. However, dynamically inferred
masses in the literature vary widely, ranging from $0.5\times10^{12}$
to $2.5\times10^{12}\msun$ across different studies~(e.g., \citealp{Wilkinson99,Xue08,Gnedin10,Gibbons14,WE15b}; see~\citealp{Wang} for a recent compilation of measurements). At least part of the
discrepancy originates from the different assumptions involved in
different methods. Hence, it is interesting to investigate the
intrinsic accuracy of a generic dynamical method that makes minimal
assumptions, which 
could then be interpreted as a lower limit on the
systematic uncertainty in dynamical mass estimation. Such a study
is also timely given the huge amount of phase-space data for stars in
the Galaxy being obtained by a new generation of instruments such as
GAIA~\citep{GAIA}.

To this end, we apply our generic dynamical method to five haloes
from the Aquarius simulations, a set of cosmological zoom-in
simulations of the formation and evolution of MW sized haloes in the
$\Lambda$CDM cosmology~\citep{Aquarius}. We fit for the mass and
concentration parameters of each halo using both the dark matter
particles, and the ``halo stars'' from the particle tagging method of
\citet{Andrew} as tracers. We find that while the dark matter (DM) tracers
recover the halo parameters accurately, the tagged stars result in
$\sim20\%$ bias in the dynamically fitted parameters. We give a brief
review of the oPDF method in Section~\ref{sec:Method}. The
applications to DM and stars are presented in Sections~\ref{sec:DM}
and \ref{sec:star}, with the data described in Section~\ref{sec:data}
and a discussion on the half-mass constraint in
Section~\ref{sec:Mhalf}. We summarize the results and conclude in
Section~\ref{sec:summary}. 

\section{The oPDF method}\label{sec:Method}
Below we briefly review the oPDF method developed in Paper~I. A likelihood estimator and a non-parametric profile reconstruction method were developed in Paper~I which show similar efficiency in making use of the dynamical information. We restrict our attention to the likelihood method throughout this paper.

\subsection{The oPDF}
In a steady-state system, phase space continuity implies a fundamental
DF, 
\begin{equation}\label{eq:oPDF}
 dP(\lambda|{\rm orbit})/d\lambda \propto dt(\lambda|{\rm orbit})/d\lambda,
\end{equation} where $\lambda$ 
is an affine parameter specifying the position of a particle on a
given orbit. That is, for any given orbit, the probability of
observing a particle at a given position $\lambda$ is
proportional to the time it spends at that position. In a spherical
potential, the orbits of particles are described by their conserved
binding energy, $E=-\left(\frac{1}{2}(\vt^2+\vr^2)+\psi(r)\right)$ and
conserved angular momentum, $L=r\vt$, where $\vr$ and $\vt$ are the
radial and tangential velocities, and $\psi(r)$ is the potential at
radius $r$. Taking $r$ as the affine parameter,
Equation~\eqref{eq:oPDF} becomes,
\begin{equation}
 dP(r|E,L)=\frac{dt}{\int dt}=\frac{1}{T}\frac{dr}{|\vr|},
\end{equation} 
where $T=\int_{r_{\rm p}}^{r_{\rm a}} dr/|\vr|$ is the period of half
an orbit, with $r_{\rm p}$ and $r_{\rm a}$ being the peri- and
apo-centre radii of the orbit. When radial cuts ($r_{\rm min}, r_{\rm
max}$) are imposed, we only need to replace the orbital limits,
$r_\mathrm{a}$, with $\min(r_\mathrm{a}, r_{\rm max})$ and
$r_\mathrm{p}$ with $\max(r_\mathrm{p}, r_{\rm min})$, since
Equation~\ref{eq:oPDF} holds within any radial range. Taking the
radial action angle, $\theta$, which we call the phase angle, as the
affine parameter, the oPDF becomes a uniform distribution,
\begin{equation}\label{eq:phase}
 dP(\theta|E,L)=d\theta,
\end{equation} where
\begin{equation}\label{eq:theta}
\theta(r)=\frac{1}{T}\int_{r_{\rm p}}^r \frac{dr}{\vr}.
\end{equation} This uniform distribution with $\theta \in [0,1]$ is also known as the random phase principle or orbital roulette~\citep{Roulette}.

\subsection{Uniform phase diagnostics}

For a steady state tracer, if one defines a normalized mean phase
 deviation~\citep{Roulette} by
\begin{equation}
 \bar{\Theta}=\sqrt{12N}(\bar{\theta}-0.5), \label{eq:meanphase}
\end{equation}
then when the sample size, $N$, is large enough the uniform phase
distribution of $\theta$ should result in $\bar{\Theta}$ being
distributed like a standard normal variable.  Hence, for a real
sample, $\bar{\Theta}^2$ can be used as a measure of the difference of
the actual phase distribution from the expected uniform distribution.

\subsection{The radial likelihood estimator}
Given a tracer and an assumed potential, one can predict the expected
radial PDF of each tracer particle using
\begin{equation}
 \D P(r)=\frac{1}{N}\sum_{j=1}^N \D P(r|E_j,L_j),
\end{equation} 
where $E_j$ and $L_j$ are the energy and angular momentum of particle
$j$ under the assumed potential. If we bin the data radially into $m$
bins, the expected number of particles in the $i$-th bin is given by
\begin{equation}
\hat{n}_i= N\int_{r_{\mathrm{l},i}}^{r_{\mathrm{u},i}} \frac{\D P(r)}{\D r} d r,
\end{equation}
where $r_{\mathrm{l},i}$ and $r_{\mathrm{u},i}$ are the lower and
upper bin edges. The binned radial likelihood is given by:
\begin{align}
 {\cal L}&=\prod_{i=1}^{m} \hat{n}_i^{n_i} \exp(-\hat{n}_i)\\
 &=\exp(-N) \prod_{i=1}^{m} \hat{n}_i^{n_i},
\end{align}
where $n_i$ is the observed number of particles in the $i$-th bin. The best-fitting potential is defined to be the one that maximizes this likelihood.

\section{Data}\label{sec:data}
We use the Aquarius simulations~\citep{Aquarius}, a set of
cosmological zoom-in simulations of the formation and evolution of
MW-sized haloes, for this analysis. The five simulations we use (labelled ``A'' to ``E'') were run at a series of resolutions and we only use the second highest resolution (level-2) runs, which have a particle mass of $\sim10^4\msun$ so that each halo is resolved with $\sim 10^8$ particles. We consider two types of tracers
in the halo: DM particles and star particles. Because Aquarius is a
DM-only simulation, the star particles are a subset of DM particles
selected with a particle tagging technique~\citep{Andrew}.

The oPDF method laid out above assumes a steady-state system with a
spherical potential. The real halo may deviate from these assumptions
in many respects, for example, by being aspherical, evolving or having
substructures. \hl{We expect these deviations to bias the fit,
and our aim is to quantify these systematic errors. To this end, we will
use a large sample to ensure that the statistical noise as inferred
from the likelihood estimator is much smaller than the level of
accuracy of interest.} Using Monte Carlo realizations we found
in Paper~I that the typical error in halo profile parameters is
$0.1/\sqrt{N/1000}$~dex for $N$
particles. Wherever possible, we will use samples with $N\sim 10^6$
particles leading to statistical errors of the order of only $\sim 1$ percent 
in the dynamically derived parameters, the mass, $M$, and the concentration, $c$.

\subsection{DM Samples}
For each halo, we create a tracer of the DM consisting of $10^6$ randomly sampled DM particles. 
To constrain the potential profile of a halo all the way out to the
virial radius, we adopt an outer cut of $300$~kpc, which is slightly
larger than the virial radius of the Aquarius haloes ($200$ to
$250$~kpc). 
We also adopt an inner cut of $1$~kpc, chosen to avoid
convergence issues~\citep{Navarro}, and to
suppress the effect of any ambiguity in the definition of the
centre of a real halo. By default, we use all the particles within the
above radial range, no matter whether the particle belongs to the
Friends-of-Friends halo or not. The Hubble flow is ignored throughout
this analysis, since the scale at which it becomes important is given by $GM/R\sim (HR)^2/2$, yielding $R \sim 1~\mathrm{Mpc}/h$
for a MW sized halo.\footnote{We have explicitly checked that including the Hubble flow
produces little difference in our results.}  



\subsection{Tagged Star Samples}
In reality one does not, of course, observe dark matter directly. A realistic
tracer population would be the stars in the halo of a galaxy. In this
section we apply the oPDF method to the Aquarius stellar
haloes calculated by \citet{Andrew}. These stars are identified in the
output of the dark matter only simulation by tagging dark matter
particles over time following the star formation history given by 
the \textsc{GALFORM} semi-analytical model
of galaxy formation~\citep{Cole94,Cole2000,Bower06}. The dynamics of the stars are then identical to
the dynamics of the tagged dark matter particles. Since the dark
matter particles are dissipationless, this tagging method does not
resolve stellar discs. Nor does it take into account the effects of baryon dissipation on 
the gravity of the system. As a result, the distribution of stars in the inner galaxy is not quite
realistic. Despite this limitation, the particle-tagging method
provides a realistic model for the stripping and distribution of
accreted stars in the \hl{simulated} outer halo, since the accreted stars 
follow the same collisionless dynamics as the dark matter particles on
large scales~\citep[see][for a controlled comparison of particle-tagging to hydrodynamical simulations]{LeBret15}. 
\hl{Recently, \citet{Cooper13} have applied this technique to large-scale cosmological simulations and have shown that
it produces galactic surface brightness profiles that agree well with the outer regions of stacked galaxy profiles from SDSS.}

To test the oPDF method with a realistic tracer population, for each
halo we  use the accreted stars from the particle-tagging
technique. In addition, we exclude particles inside 10~kpc of each halo
as the presence of a disc in a real galaxy violates the spherical symmetry assumption for the potential, and because the
lack of such a disc in the simulated halo makes the mock data
less realistic at small radii. As with the dark matter tracers, an
outer radius cut of 300~kpc is applied to each halo. In a forthcoming
paper (Wang et al., in prep), we will extend this study to a larger
sample of Local Group haloes in which the stars are taken from
hydrodynamical simulations.

Due to the limited resolution of the dark matter simulation, each
tagged particle may represent many stars with varied stellar masses,
and one dark matter particle may be tagged multiple times representing
stars formed at different epochs. However, the dark matter particles
in the original simulation are followed dynamically without knowledge
of the stellar mass weighting or multiple tagging. Hence the dynamics of
these tracers are only resolved to the level of the tagged dark matter
particles.\footnote{From a statistical point of view, the weighted
distribution contributes an additional uncertainty to the stellar mass of each
particle, making the star particle counts in bins a Compound Poisson
process rather than a Poisson process. So strictly speaking, the
current likelihood model does not apply to the weighted distribution.}
For the purpose of dynamical modelling, we mainly use the unique set
of tagged particles without any stellar mass weighting. This
leaves us with $5$ -- $8\times 10^{5}$ unique tagged particles for each
halo in the level-2 simulations. In the following, we continue to use
the term stars to refer to these unique sets of tagged particles. 

\begin{figure}
 \includegraphics[width=0.5\textwidth]{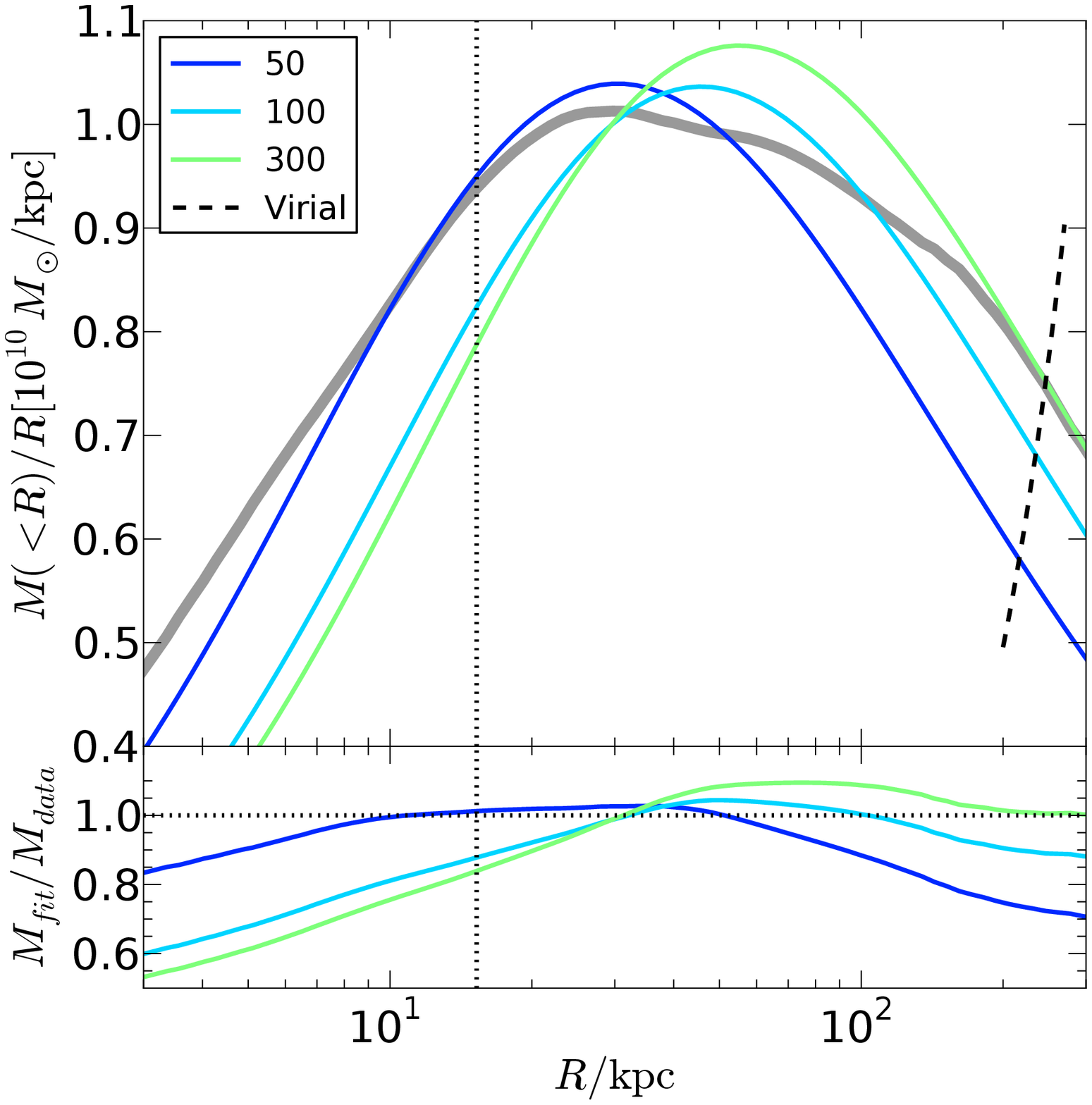}
 \caption{The mass profile (scaled as $M(<R)/R$) of halo~A. In the
 upper panel, the grey shaded line shows the true mass profile
 of the dark matter distribution, while the different coloured
 lines show NFW profiles from maximum likelihood fits within $50,100$
 and~$300$~kpc respectively. 
 The black dashed line (labelled virial) shows Eq.~\eqref{eq:vir}. It crosses
each coloured line at the virial radius of each profile. The vertical dotted line marks the scale radius, $r_{\rm s}$.
The lower panel shows the ratio of the fitted mass and the true mass as a function of the enclosing
 radius.}\label{fig:A4Mass}
\end{figure}

\subsection{Template profiles: defining the true potential and halo
  parameters}\label{sec:template}

To fit the halo potential using the oPDF method and assess any biases
in the fit, we need to parametrize the potential with some functional
form and also define the true parameters of the potential function.

One choice of parametrization is the widely used NFW profile~\citep{NFW96,NFW97},
\begin{equation}
 \rho(r)=\frac{\rho_{\rm s}}{(r/r_{\rm s})(1+r/r_{\rm s})^2},\label{eq:NFW}
\end{equation}
where $r_{\rm s}$ is the scale radius at which $d\ln \rho/d\ln r=-2$ and
$\rho_{\rm s}$ sets the density at this radius.
These two parameters can be
analytically related to the virial mass and concentration
parameters. The virial mass is defined as the mass inside a virial
radius, $R_{\rm v}$, where the enclosed density is $\Delta_{\rm v}$
times the critical density of the universe
\begin{equation}\label{eq:vir}
 M=\frac{4\pi}{3}\Delta_{\rm v} \rho_{\rm c} R_{\rm v}^3.
\end{equation} 
Throughout this paper, we adopt $\Delta_{\rm v}=200$. The
concentration parameter is defined as $c=R_{\rm v}/r_s$. With
this parametrization, one may choose to use the best-fitting
parameters of the density profile as the true parameters. However,
such a choice would be problematic if an NFW profile is not a good description
of the halo density profile in question, in which case the best-fitting NFW
parameters could depend on how the fit is performed. To
demonstrate this, we fit the density profile of halo~A using a maximum
likelihood method. Note that dynamical modelling is not involved here,
and the fit is purely to characterize the true mass
distribution of the halo. The extended likelihood \citep{Barlow90},
${\cal L}$, can be written as
\begin{equation}
 \ln {\cal L}=\sum_i \ln \rho(r_i)-N_{\rm pred},
\end{equation} 
where $N_{\rm pred}=\int_{\rm window} \rho(r)/m_{\rm p}\,\D^3r$ is the
predicted number of particles in the data window, with $m_p$ being the
particle mass, and $\rho(r)$ the NFW density profile given by
Eq.~\eqref{eq:NFW} with parameters $(\rho_{\rm s}, r_{\rm s})$. $r_i$ is the radial
coordinate of the $i$-th particle and the summation runs over all the particles in the data window.
This method is, in the limit of infinitesimal bins, equivalent to fitting
to a binned profile provided one takes account of the Poisson
distribution of the counts inside each bin.
We fit the dark matter distribution around
halo~A  over several different radial ranges, with outer cuts of
$50,100$ and~$300$~kpc respectively. The best-fitting mass profiles along
with the real mass profile are shown in Fig.~\ref{fig:A4Mass}.
It is obvious that the fits differ from each other, and none of them
describes well the full mass profile out to the virial radius. The 
inferred virial masses can differ by more than $30\%$.
We note that halo~A is an extreme example which deviates
grossly from NFW, while the remaining four Aquarius haloes agree much
better with the NFW form.

Given the poor performance of the NFW parametrization for halo A, it
would be problematic to define the true halo mass, concentration or
potential parameters from a best-fitting NFW profile.  Put another
way, any fit that adopts an NFW parametrization also suffers from
systematics introduced by deviations of the real halo profile from
the NFW form. To eliminate this systematic uncertainty, we will
describe the potential using parametrized template profiles that are
able fully to match the true profile. For each halo, we first extract
the true potential profile from the spherically averaged density
profile. Specifically, the potential at a given point is evaluated as
\begin{equation}
 -\psi(r)=G\sum_{r_i<r}\frac{m_i}{r}+G\sum_{r_i\geq r} \frac{m_i}{r_i},\label{eq:PotEst}
\end{equation} 
where $r_i$ and $m_i$ are the radial position and mass of the $i$-th
particle. In practice, the profile is extracted at a sequence of radii
and then interpolated at any other radius.
Once a true profile is extracted, we generalize it to a two parameter
family by varying its scale and amplitude. Specifically, for each real
profile, $\psi(r)=f(r)$, we generate a parametric template as
\begin{equation}
\psi(r)=A f\left(\frac{r}{B}\right),\label{eq:template}
\end{equation} 
where $A$ and $B$ are dimensionless scale parameters. These two
parameters can be mapped to $M$ and $c$ following the procedure in
Appendix~\ref{app:template}. The true parameters ($M_0,c_0$) of the
halo are unambiguously \emph{defined} by locating where in the true
density profile the spherical overdensity matches the virial overdensity
criterion and where the profile has a logarithmic slope of $-2$.

We will consider both the NFW and the template parametrizations when
fitting the potential.

\section{Application to DM Haloes}\label{sec:DM}
\subsection{The dynamical state of Aquarius haloes}

\begin{figure*}
 \includegraphics[width=\textwidth]{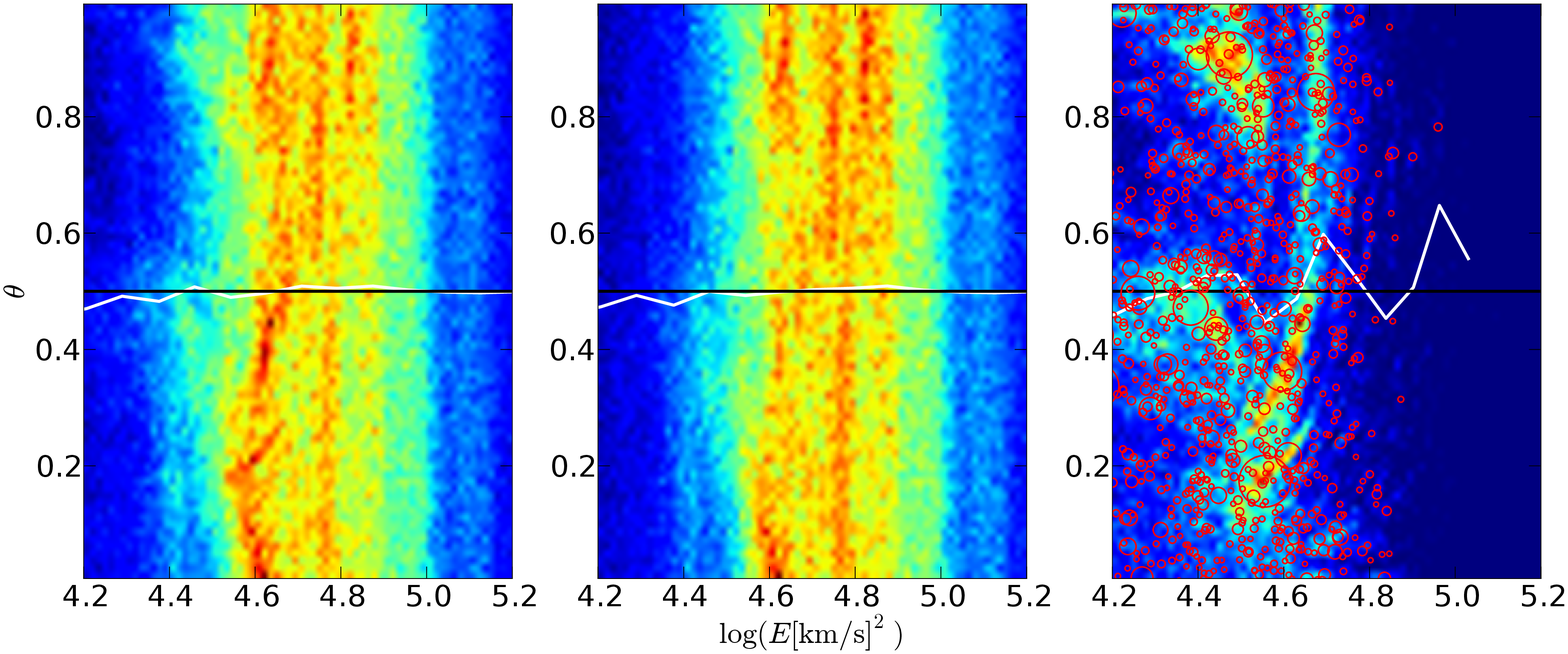}\\
 \includegraphics[width=\textwidth]{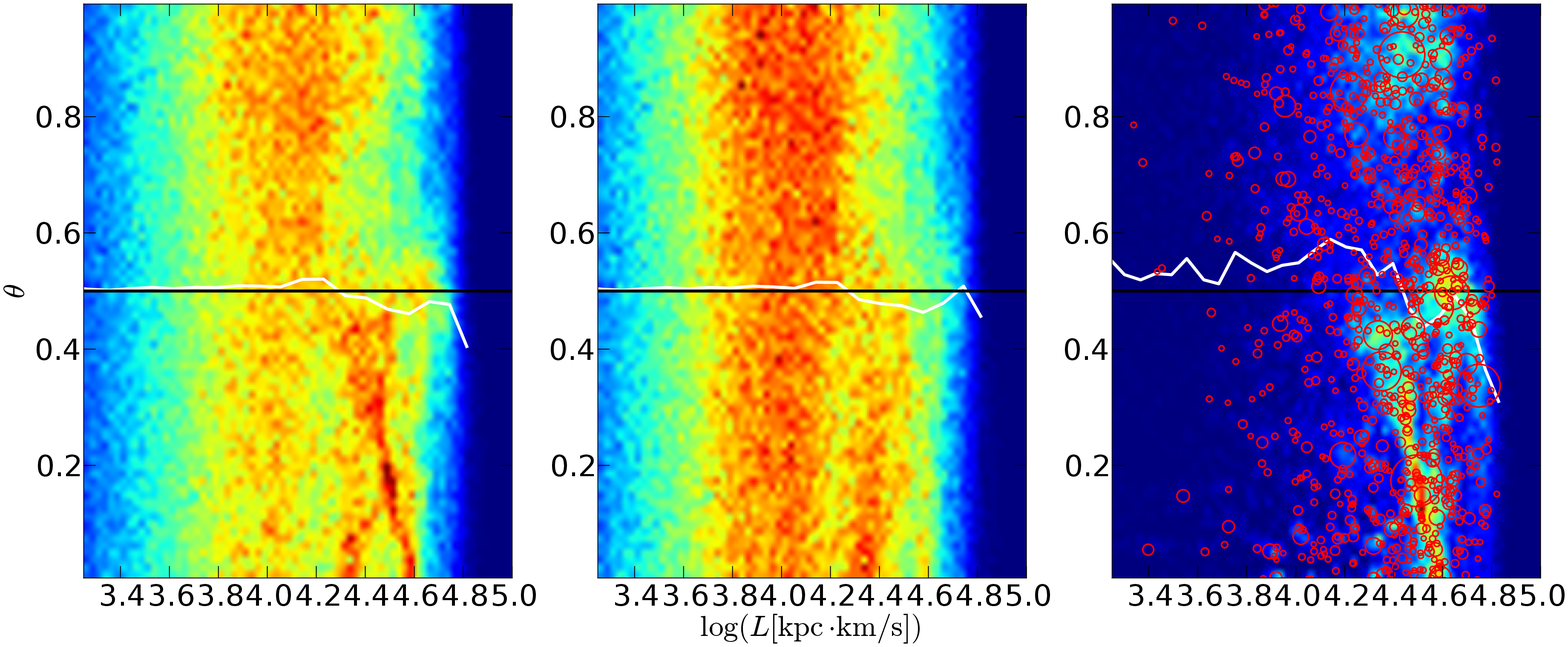}
 \caption{The phase space distribution of particles in halo A. Top and
 bottom panels show the distributions in $E-\theta$ and $L-\theta$
 spaces respectively. The left column shows the distributions of all
 the particles in the sample. The middle column shows that with
 subhalo particles removed. The right column shows the distributions of
 subhalo particles alone. Each subhalo with more than 1000 particles
 is also marked by a red circle in the right hand panels. In all the
 panels, only particles with $1<r<300~{\rm kpc}$ are
 used. The white lines are the median $\theta$s. 
 The images colour code the number of particles in
 each pixel. The contrast of each
 panel has been individually optimized.}\label{fig:PhaseImage}
\end{figure*}

Once the real potential is known (Eq.~\ref{eq:PotEst}), we can
examine the distribution of particles in ($\theta, E, L$) space prior
to any fit. According to Eq.~\eqref{eq:phase}, for any system in a
steady state, $\theta$ should be uniformly distributed for particles in
any bin of $E$ or $L$.  In Fig.~\ref{fig:PhaseImage}, we show the
example of halo A in such coordinates.  In the left panels, all the
particles within $1-300$~kpc from the halo centre are used. We are not
concerned with the distributions along the $E$ and $L$
directions. Along the $\theta$ direction, overall, at fixed $E$ or $L$ the particle
distributions are close to uniform. However, one can still identify
clumps in phase space which perturb the uniformity. In the rightmost
panels only particles from subhaloes identified by {\sc
subfind}~\citep{SUBFIND} are plotted. The coordinates of subhaloes
with more than 1000 particles are overplotted as red circles, with
larger circles corresponding to more massive subhaloes. The remaining
particles representing a smooth component are plotted in the middle
column.  Comparing the three columns, it is obvious that substructures
introduce perturbations to the uniform $\theta$-space distribution of
the host halo. These perturbations are twofold: first the particles
inside subhaloes are locally clustered and break the uniform
distribution; secondly, the potential of the subhaloes exists as
perturbations to the potential of the smooth host halo, affecting the
orbits of nearby particles. The existence of locally clustered
structures makes the real particle distribution noisier than a Poisson
realization of a smooth uniform field, and degrades the consistency
between the two. In principle, substructures can be defined as locally
overdense structures in phase space, and a phase space substructure
finder could be designed to excise them and optimize the uniformity of
the distribution of the remaining background particles.  In practice,
as we find subhaloes with {\sc subfind}, the removal of substructures may not
always increase the dynamical uniformity of the system unless the
substructure finder is designed to do so.


Compared with the smooth component, subhaloes occupy relatively low
binding energy and high angular momentum orbits. Despite the clumpy
distribution of subhalo particles, they do not appear to
bias the $\theta$ distribution significantly in any particular direction. We
will come back to this point when fitting the mass and concentration
of the haloes. 

\begin{figure*}
\myplottwo{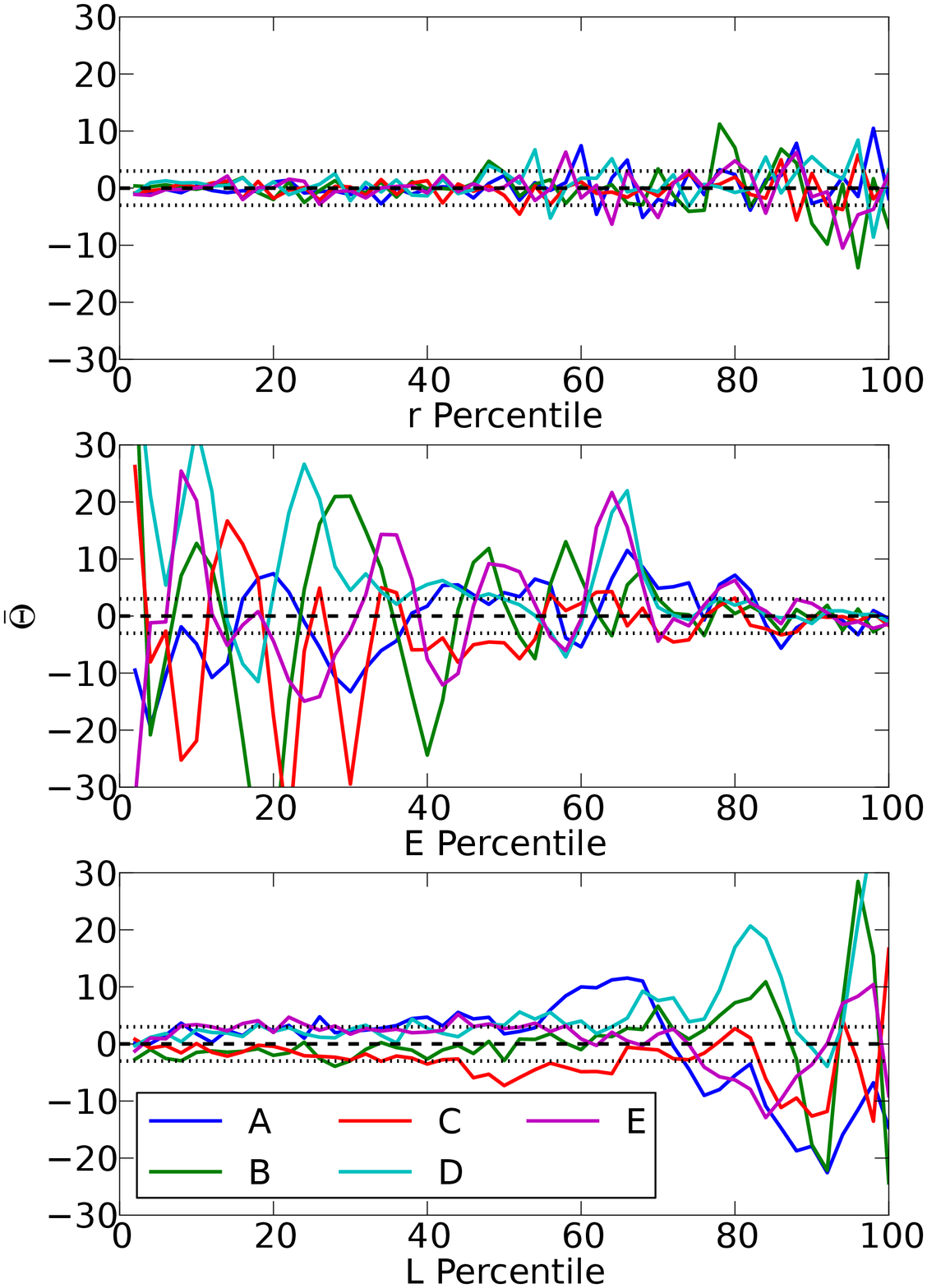}{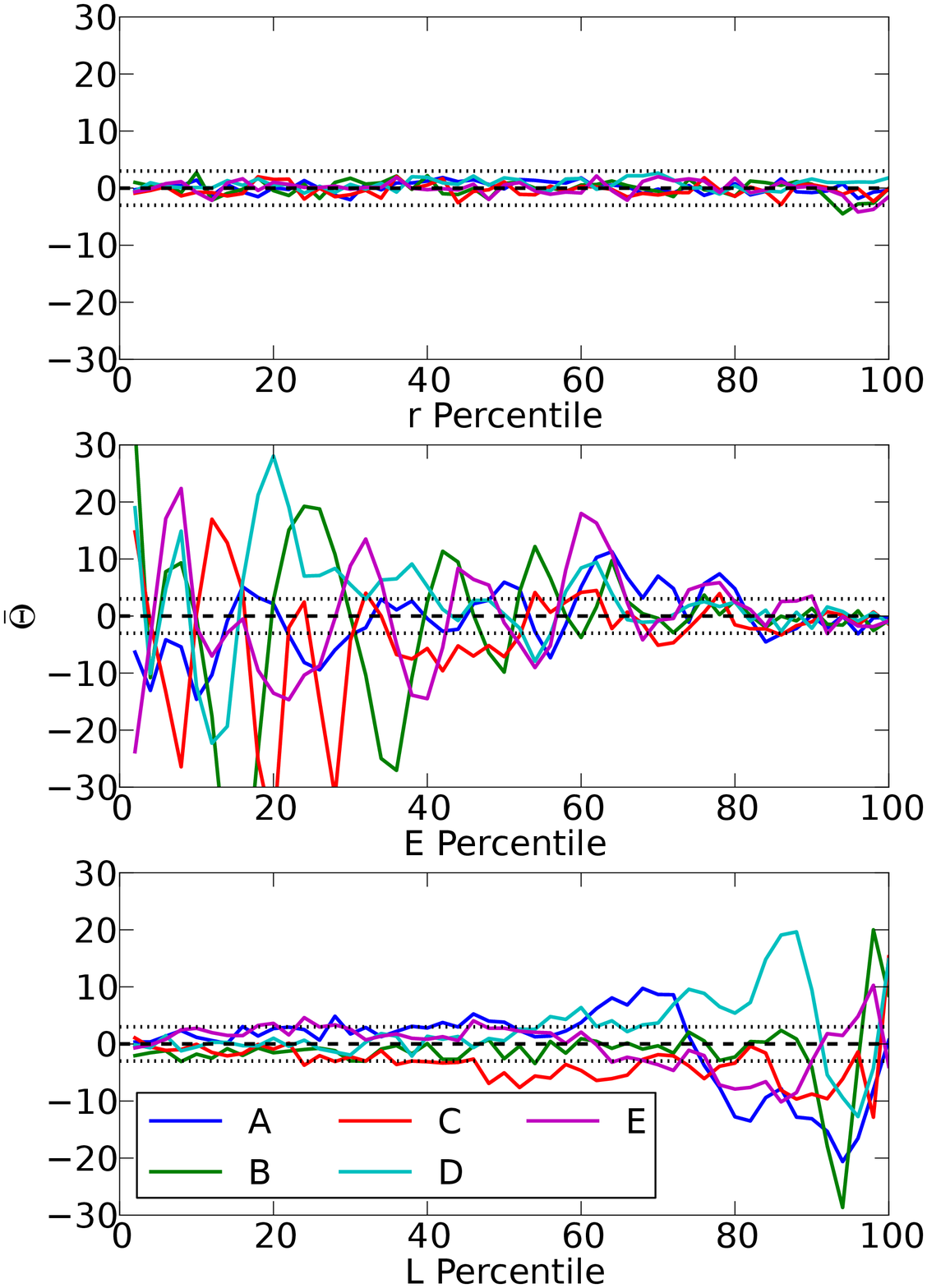}
\caption{The normalized mean phase deviation profile of Aquarius DM
haloes. From top to bottom, we bin the halo particles according to
their $r,E,L$ coordinates respectively, with equal numbers of particles
in each bin. The mean phase deviation, $\bar{\Theta}$, is evaluated
in each bin, and plotted as a function of the percentile values in
the respective coordinate. Different colour lines represent different
haloes. The dashed and dotted reference lines mark the 0 and $\pm
3\sigma$ discrepancy levels. The left panels show the profiles of the
full sample. The right panels show that with subhalo particles
removed. \cl{See the online version for a coloured plot.}}\label{fig:TSprofDM}
\end{figure*}

In Fig.~\ref{fig:TSprofDM}, we explore deviations from a steady state of the DM
tracer at different values of $r,E$ and~$L$ in terms of the normalized
mean phase deviation, $\bar{\Theta}$, which measures the discrepancy
level from a uniform distribution. For each halo, we calculate the
mean phase within bins of phase-space coordinates $r$, $E$ or~$L$.  We
create the bins with equal numbers of particles per bin, so that they
have the same statistical noise, allowing direct comparison of
$\bar{\Theta}$ across the bins.  The bins are labelled by the
percentiles of the respective sorted phasespace coordinate, $r$, $E$
or~$L$.  Recall that, if the tracer is in a steady state, then in the large
sample limit, $\bar{\Theta}$ is distributed like a standard normal
variable.  

Consistently with the physical picture displayed in
Fig.~\ref{fig:PhaseImage}, the DM particles have a mean phase
deviation broadly consistent with zero. As seen from the left column,
the discrepancy is most significant at large radius, low binding
energy and high angular momentum, revealing a higher level of
systematics at these locations. Note low $E$ and high $L$ regions are
also where subhaloes are most abundant as seen in
Fig.~\ref{fig:PhaseImage}, and it is also well known that subhaloes tend to
occupy the outer halo \citep[see, e.g.,][]{Aquarius}. The panels of
the right hand column are the same as the left, but with subhalo
particles removed from the tracer. In calculating the radial profile,
the radial limits of the data window, $r_{\rm min}$ and $r_{\rm max}$,
have been adjusted to the bin edges, so the radial profile examines
the local uniformity of particles. After removing subhalo particles,
local dynamical consistency is significantly improved at large
radius. However, we see from the middle and bottom panels of
Fig.~\ref{fig:TSprofDM} that this has little effect on the dynamical
consistency within individual orbits over the full radial range.


Note that $\bar{\Theta}$ correlates with the depth of the proposed
potential and a positive $\bar{\Theta}$ indicates the current
potential is deeper than a best-fitting potential (Paper~I). As seen
in Fig.~\ref{fig:TSprofDM}, at large $r$, $L$ and low $E$, the mean
phase deviation can be significantly higher than one would expect from
a uniform distribution, which would lead to a level of systematic
uncertainty significantly larger than the statistical noise in the
best-fitting potential. However, overall the fluctuation is still
stochastic with no preferred sign. This indicates that if one is
going to fit the potential, then deviations from our model assumptions
are unlikely to bias the model parameters in a particular direction; 
instead, the biases would fluctuate stochastically. Despite this, these biases 
are still systematic rather than statistical in nature, as they are tied to
the model assumptions, not to the sample size.
In the following section, we aim to quantify the level of such
systematic uncertainty in the best-fitting parameters of the halo
potential.

\begin{figure}
 \includegraphics[width=0.5\textwidth]{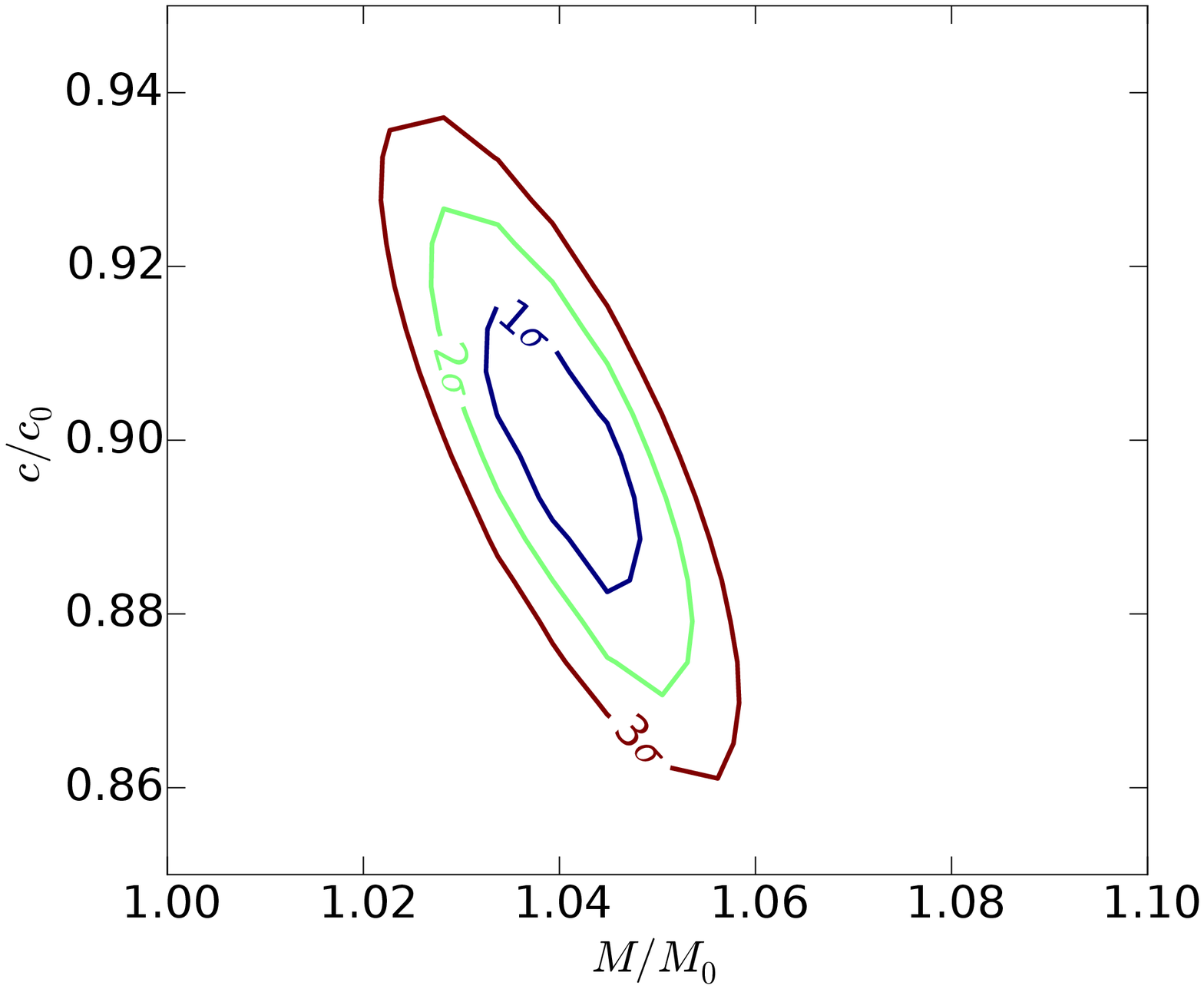}
 \caption{The 1, 2 and 3$\sigma$ confidence contours for the full sample of
 halo A fitted with the oPDF likelihood using the template
 profile. The parameters are in units of their true
 values.}\label{fig:A2err}
\end{figure}

\subsection{Fitting the halo potential with DM as tracers}

\begin{figure*}
 \myplottwo{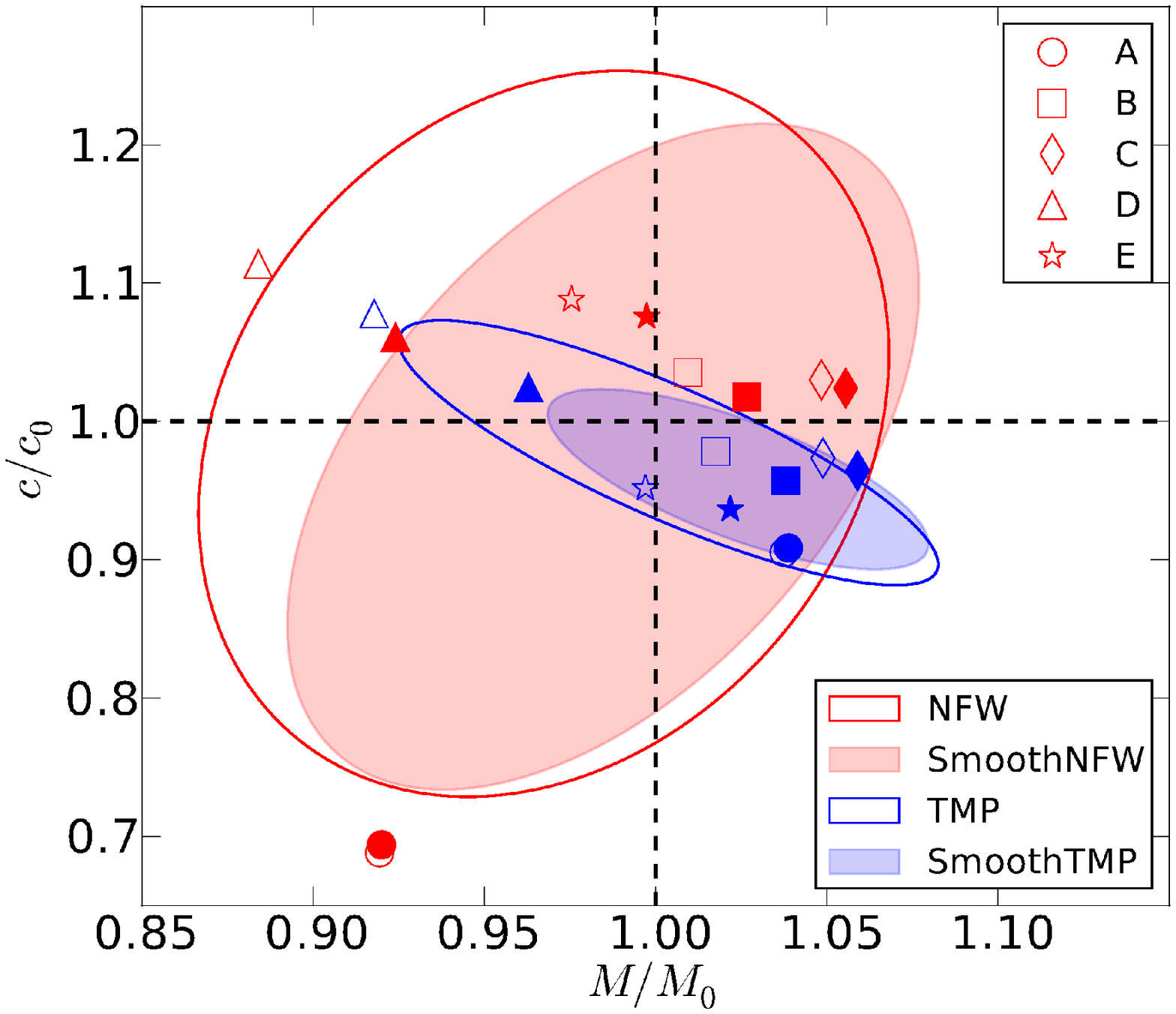}{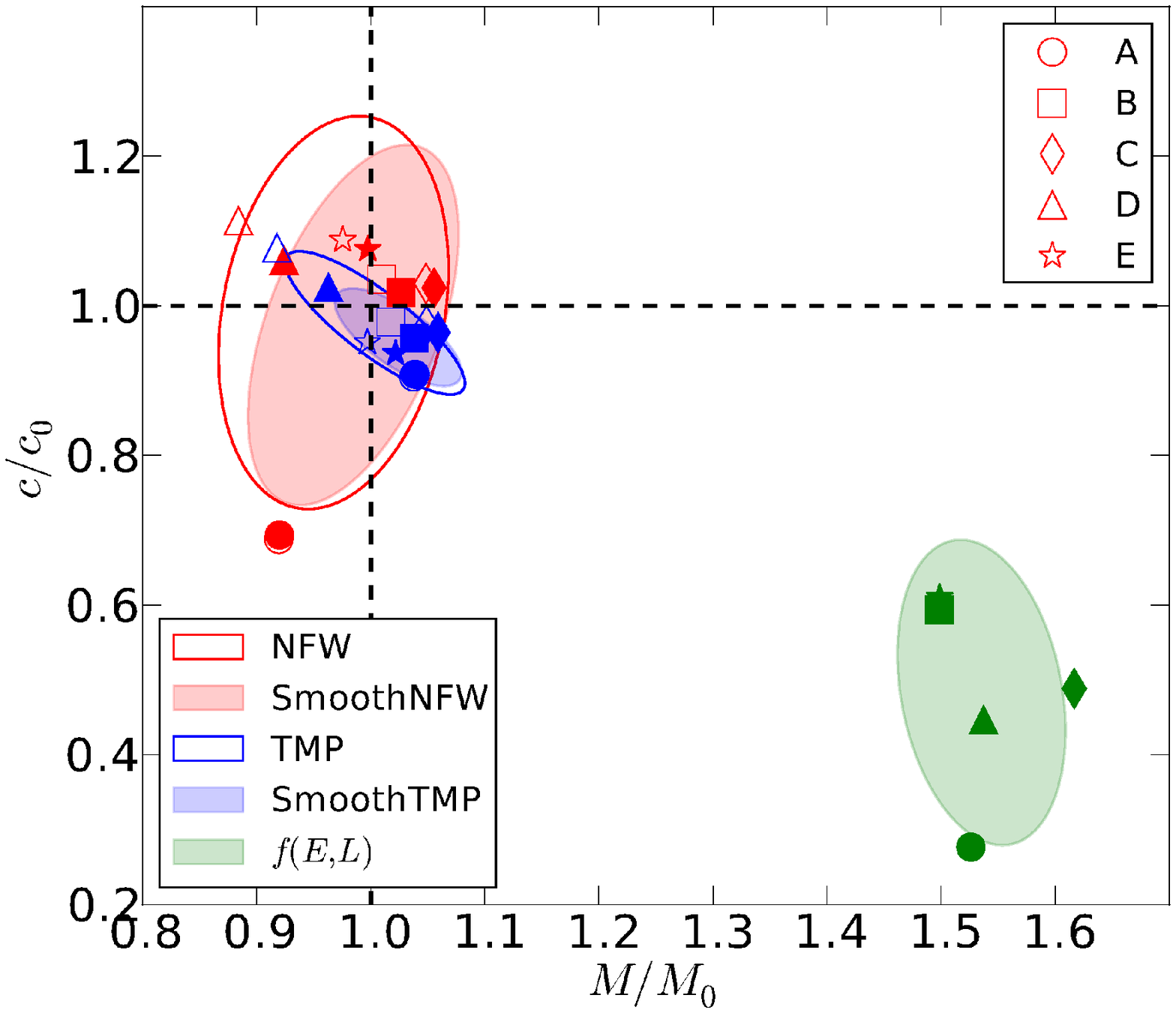}
 \caption{Left: fitted parameters of Aquarius haloes using the radial
 likelihood estimator. Different shaped symbols denote different haloes. The
 red and blue colours denote the fitting results using NFW and
 template profiles (labelled TMP) respectively. In both cases, the open symbols show
 the fits with the full samples, while the filled ones show those for
 the smooth samples, i.e., with subhalo
 particles excluded. For each combination of sample and profile, we combine the
 five haloes to estimate a mean and a covariance matrix for the
 parameters, and plot the $1\sigma$ contour (the ellipses, open or filled) in the same
 style for a bivariate Gaussian with the estimated mean and
 covariance. Right: same as the left, but also showing the fits from
 \citet{Wang} to the  smooth DM sample using \hl{a $f(E,L)=L^{-2\beta}F(E)$ model} (green
 symbols and ellipse). \cl{See the online version for a coloured plot.}}\label{fig:DMFit}
\end{figure*}

With the potential functions and their true parameters defined in
Section~\ref{sec:template}, we can proceed to fit the potential
profiles with our oPDF method, and quantify the level of systematic
uncertainty in the fitted parameters. We adopt the binned radial
likelihood estimator, with 50 logarithmic bins. Fitting with a
different number of radial bins gives consistent
results.\footnote{Adopting the Anderson-Darling estimator described in
Paper~I~\citep[see also][]{Roulette} increases the parameter scatter
to $\sim20\%$, due to its poorer accuracy.} For both NFW and template
parametrizations, we fit two datasets: 1) all the dark matter
particles inside $1-300$~kpc, i.e, the full sample; 2) the former but
with all the subhalo particles removed, i.e, the smooth
sample. Because we aim to quantify the systematic uncertainties due to
deviations from model assumptions, we need to make sure that the
statistical noise, which is determined by sample size, is small
enough. As an example, in Fig.~\ref{fig:A2err} we show the statistical
confidence contours of halo~A from the template fit. These error
estimates are consistent with the scatter among independent subsamples
of the parent halo. The $1\sigma$ error is around $0.005$~dex for our
sample of $10^6$ particles, quite consistent with our expected scaling
of $0.1/\sqrt{N/1000}$~dex. Such an accuracy should be sufficient for
detection of systematic biases larger than $1\%$.

The best-fitting parameters in units of the true parameters are
plotted in the left panel of Fig.~\ref{fig:DMFit}. Overall, the fitted
$(M,c)$ parameters largely agree with their true values, with a bias
generally smaller than $10\%$. The typical bias quantified by the
scatter among the five haloes is $\sim5\%$ as listed in
Table~\ref{table:summary}. For each parametrization and dataset, we
combine the five haloes to estimate a mean and a covariance matrix for
the parameters, and plot the one-sigma contour for a bivariate
Gaussian with the estimated mean and covariance. Note these contours
are an estimate of the systematic uncertainties, since the statistical
noise of the model is negligible given the sample sizes. Consistently
with our expectation from the mean phase profiles, there is not a
definitive systematic bias but rather, as far as we can tell from the
small sample of haloes, the scatter is mostly stochastic from halo to
halo.

\begin{figure}
\myplot{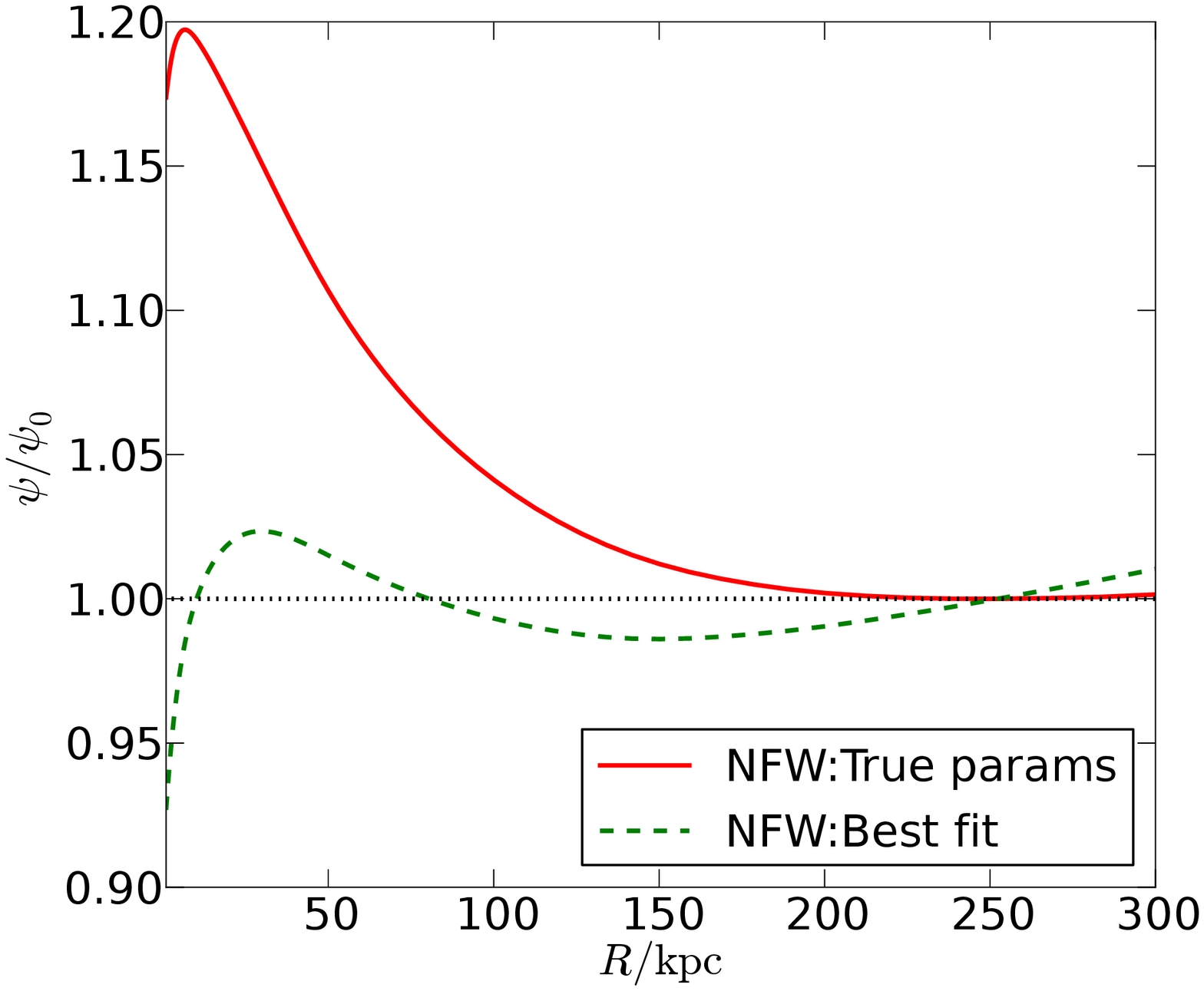}
\caption{The potential profile of halo~A and its NFW
parametrization. We plot the ratio between the parametrized NFW
potential, $\psi$, and the true potential, $\psi_0$, of the halo as a
function of radius.  For the dynamical modelling only the potential
difference is relevant and so the zero-points of the NFW potentials have
been adjusted to produce the true potential value at the virial
radius. The red solid line corresponds to the NFW potential profile
using the true parameters and the green dashed line to the NFW parameters
found from the oPDF likelihood estimated from all the dark matter
particles.}\label{fig:A2Pot}
\end{figure}
Comparing the fits with and without subhalo particles, there is not a
significant improvement in the latter. When NFW profiles are adopted
in the fits, the accuracy is comparable to that achieved with template
profiles in most cases. This reflects the fact that most haloes are
well described by NFW profiles. Note that the significantly larger
confidence regions in NFW fits as marked by the ellipses in
Fig.~\ref{fig:DMFit} is caused purely by halo~A, whose dynamical fit
shows a bias in concentration up to 30\%. This is due to the fact that
the density profile of halo~A differs significantly from NFW, as is
evident from Fig.~\ref{fig:A4Mass}. In Fig.~\ref{fig:A2Pot} we show
the disagreement in halo~A from a different perspective, by comparing
the NFW parametrization of the halo potential with the true potential. 
When the set of true halo parameters are used, the NFW potential is consistently overestimated
inside the halo. The dynamical fit adjusts the parameters so that the
NFW potential agrees with the true potential to within 5 percent for
most of the radial range. The best-fit NFW potential agrees better
with the true potential, reflecting the fact that the dynamical fit
largely recovers the true potential by force-fitting the NFW
parametrization, despite giving different
parameters from the true values. It is quite interesting to see that when the template
profile is adopted, halo~A does not appear to be more biased than the
other haloes, meaning that a deviation from the NFW form does
\emph{not} necessarily mean a lack of equilibrium.

In the right panel of Fig.~\ref{fig:DMFit}, we compare our fits
to those obtained from a conventional DF method that describes the
phase space density only as a function of $(E,L)$ of the
particles~\citep{Wang}. Specifically, the phase space probability is
assumed to have the form $\D P(\VR,\VV)=f(E)L^{-2\beta}\,\D^3r\,
\D^3v$, where $\beta$ is a parameter describing the velocity
anisotropy, with $f(E)$ further determined by inverting a
double-powerlaw tracer density profile inside an NFW
potential (see Eq.~12 in \citealp{Wang} for further details). \hl{This distribution function describes a family of models with constant anisotropies, while in general more flexible models can be constructed~\citep[e.g.][]{Wojtak,Posti15,WE15}.} In
contrast to the fairly unbiased fits with our method, this $f(E,L)$
method suffers from a $\sim50\%$ net bias in the parameters. As discussed
in \citet{Wang}, this can be attributed to the fact that the $f(E,L)$
DF only describes \hl{gravitationally bound} systems by construction \hl{(see also section 6.3.1 of Paper I)}, and struggles to
match the distribution of the loosely bound particles. Because our
oPDF method has no prerequisite on the distribution of orbits (hence
no prerequisite on the energy distribution), our fits show no such net
bias. On the other hand, the $f(E,L)$ fits exhibit a comparable
amount of halo to halo scatter in the parameters to ours, reflecting that
our method does capture the minimum irreducible uncertainty associated
with steady-state models.


\section{Application to mock stellar haloes}\label{sec:star}
\begin{table*}
\caption{Basic properties of the stellar haloes, within
$10-300$~kpc. $N_{\rm tot}$ is the total number of star particles,
$N_{\rm smth}/N_{\rm tot}$ is the fraction of particles in the
smooth component and $m_{\rm smth}/m_{\rm tot}$ is the mass fraction of
particles in the smooth component.}
\label{table:star}
\begin{center}
\begin{tabular}{ccccccc}
\hline
\hline Halo & $N_{\rm tot}$ & $N_{\rm smth}/N_{\rm tot}$ & $m_{\rm smth}/m_{\rm tot}$ & $D$ \\
\hline A & $5.4\times10^5$ & 0.50 & 0.34 & 9.6 \\
\hline B & $7.4\times10^5$ & 0.72 & 0.36 & 9.6 \\
\hline C & $5.3\times10^5$ & 0.70 & 0.43 & 5.9 \\
\hline D & $8.1\times10^5$ & 0.60 & 0.52 & 5.5 \\
\hline E & $5.1\times10^5$ & 0.47 & 0.18 & 9.5 \\
\hline
\end{tabular}
\end{center}
\end{table*}

\begin{figure*}
\includegraphics[width=\textwidth]{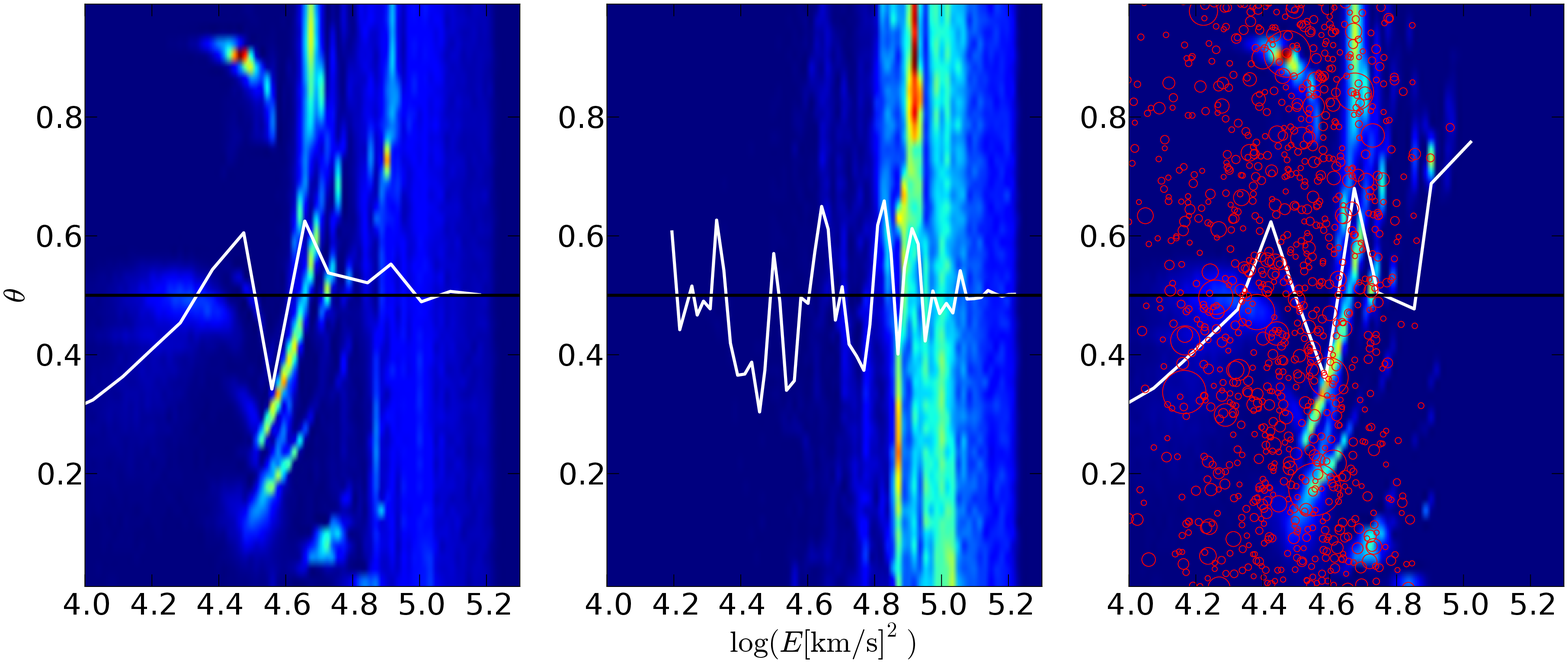}
\includegraphics[width=\textwidth]{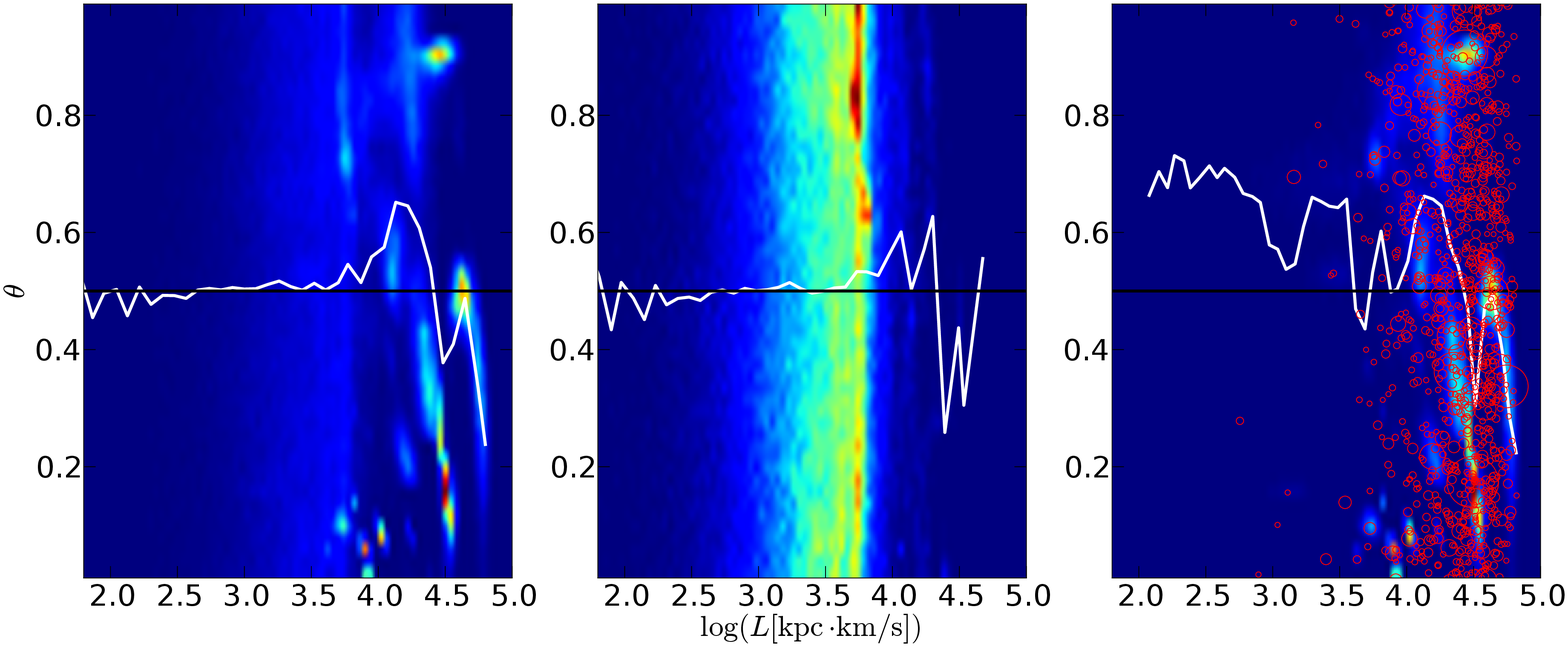}
\caption{As Fig.~\ref{fig:PhaseImage} but for the stars in the
level 2 halo~A. From left to right the distributions of all the
stars, the ``smooth'' component (all stars excluding satellites), and
those in satellite subhaloes. The contrast of each panel is individually optimized.}\label{fig:StarPhaseImage}
\end{figure*}

In Fig.~\ref{fig:StarPhaseImage} we show the phase space distribution
of stars in halo~A. Unlike the DM tracer which is only slightly
perturbed by subhaloes, the halo stars are dominated by those in the
satellite subhaloes. The mass fraction contained in satellite galaxies
is $50\%-70\%$ (Table~\ref{table:star}) in the radial range of
interest. These satellite stars are obviously not in equilibrium with
the rest, and can be observationally identified and removed as
satellite galaxies. In what follows, we will only use
the ``smooth'' component of halo stars, i.e, those excluding satellite
stars, as our tracer sample. In total, each halo has $(2-5)\times
10^5$ smooth star particles within $10-300$ kpc, yielding a statistical
uncertainty of $\sim 2\%$ in mass and concentration.

\begin{figure}
 \myplot{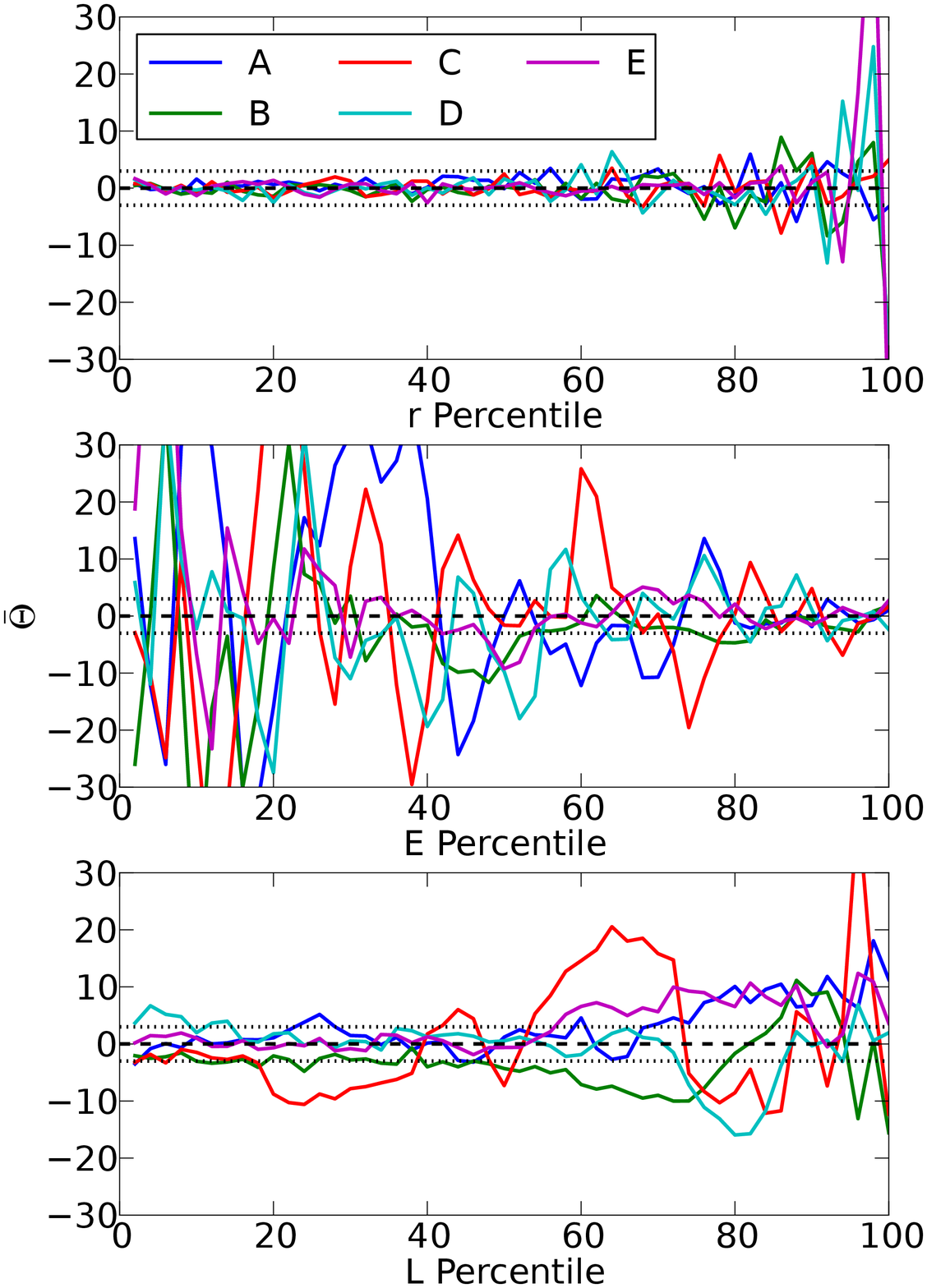}
 \caption{The mean phase deviation profile of Aquarius stellar
 haloes. This is the same as the right hand side of
 Fig.~\ref{fig:TSprofDM}, but for star particles. From top to bottom,
 we bin the star particles according to their $r,E$ and~$L$ coordinates
 respectively, with equal numbers of particles in each bin. The mean
 phase deviation, $\bar{\Theta}$, is evaluated in each bin, and
 plotted as a function of the percentile values in the respective
 coordinate. Different coloured lines represent different haloes as
 indicated in the legend. The dashed and dotted reference lines mark
 the 0 and $\pm 3\sigma$ discrepancy levels. Only the smooth component
 of the halo stars is used. \cl{See the online version for a coloured plot.}}\label{fig:TSProfStar}
\end{figure}

The mean phase deviation profile is shown in
Fig.~\ref{fig:TSProfStar}. As for the profile of DM tracers, 
overall $\bar{\Theta}$ is consistent with zero, with the highest
scatter seen at large radius, low binding energy and high angular
momentum. Note that stars with low binding energies are also those
that have been accreted recently~\citep{Wang}. The scatter in the star
profiles 
also appears higher than
that in the DM case.

\begin{figure}
 \myplot{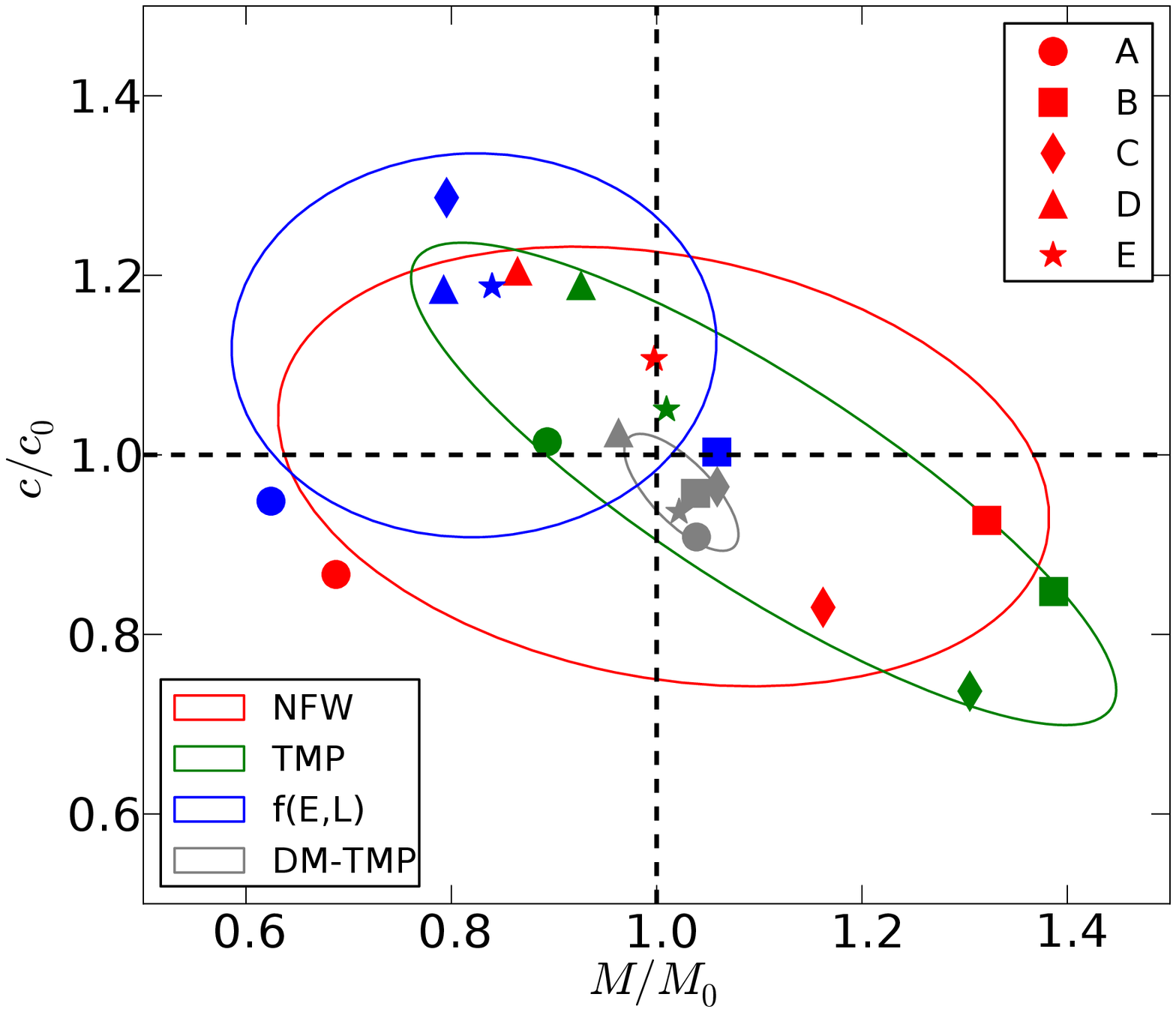}
 \caption{As Fig.~\ref{fig:DMFit}, but showing the results of
 the dynamical fits to the halo stars.  As indicated in the legends,
 symbols of different shapes represent different haloes, while
 different colours distinguish different datasets and model
 profiles. Red and green show fits to stars adopting NFW and template
 profiles respectively; blue shows the fit from \citet{Wang} to
 stars combining radial and tangential velocities using a specific $f(E,L)$ model; 
 grey shows the template fit applied to the smooth DM tracer. The symbols are the
 results of fits to individual haloes, while the large ellipses mark
 the estimated $1$-$\sigma$ confidence regions for each type (i.e.,
 combination of dataset and model) of fit estimated from the sample of
 five haloes. \cl{See the online version for a coloured plot.}}\label{fig:StarFit}
\end{figure}

The fits using stars are plotted in Fig.~\ref{fig:StarFit} for both the template and NFW profile models. For
individual haloes, the deviation from the true parameters can be as
high as 40\%.
For comparison, the fits and $1\sigma$ contours from the $f(E,L)$
method of \citet{Wang} and those from the DM tracers in the previous
section are also plotted.

Overall, we do not observe a statistically significant net bias in the
fits with the current sample of five haloes, even though the $f(E,L)$
method applied to stars is only marginally unbiased at the $1\sigma$ level. In
other words, the systematic bias varies from halo to halo in a
stochastic way. Despite this stochastic behaviour, we have checked
that the systematic bias does not change with sample size, so it is
indeed a systematic rather than a statistical error. There is a negative
correlation between the mass and concentration parameters in the
template fits, which is similar to the correlation in the statistical
noise of the two parameters. This correlation is absent in the NFW
fits only because halo A is not well described by NFW and this biases the
fit significantly. A viable explanation of this behaviour of the
systematic bias lies in the deviation of the tracer population from a
steady state.
For example, the existence of correlated phase angles in
streams and caustics implies that different tracer particles are not
independent. As a result, the constraining power of a set of particles
in a stream is less than that of an equal number of independent
particles. In the large sample limit,
when each stream is sufficiently sampled, the errors on the inferred
model parameters do not vanish but are limited by the effective
number of independent streams or particle clumps. This is an
intrinsic property of each halo. Hence it is understandable that we
are left with irreducible stochastic biases in well sampled haloes. In
addition, these residual errors are expected to exhibit similar
parameter correlations to the statistical noise.  Note that while the DM
fits exhibit only $\sim 5\%$ scatter, the scatter for the stellar
fits is typically $\sim 20\%$.

\begin{figure}
 \myplot{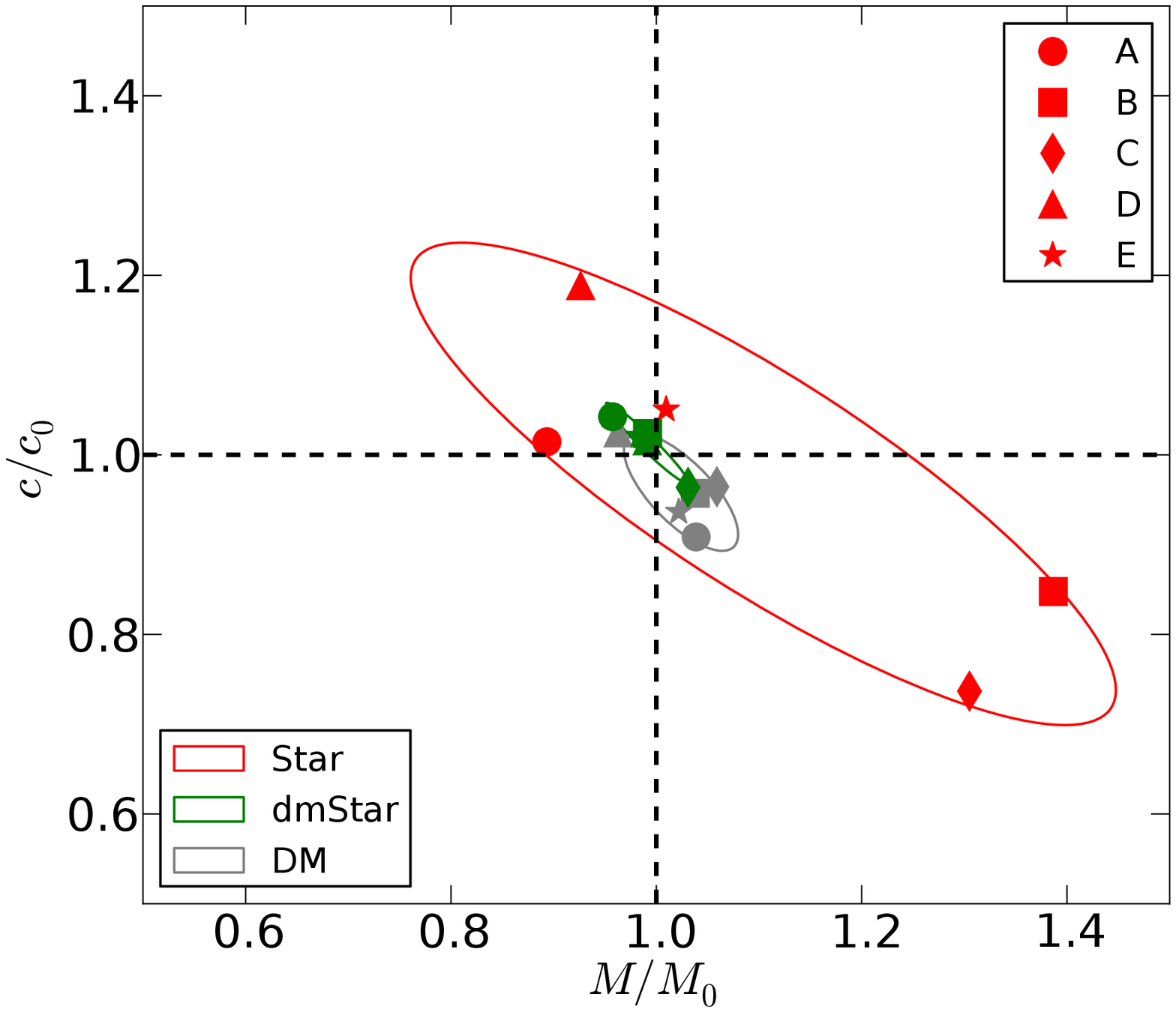}
 \caption{Comparison between the results of stellar and the DM
 dynamical fits. The ``Star'' and ``DM'' fits are the same as in
 Fig.~\ref{fig:StarFit}, using stars and DM particles
 respectively. The ``dmStar'' fit uses a sample of DM particles
 selected to have the same $E$-$L$ distribution as that of the
 stars. Subhalo particles have been removed in all three cases and
 only the template profiles are used. \cl{See the online version for a coloured plot.}}\label{fig:StarFitDM}
\end{figure}

Since stars and DM tracers have different $E$-$L$ distributions,
they occupy a statistically different set of orbits.  This implies the
tracers potentially have different spatial distributions. As a result,
they could be sampling different parts of the halo, or the same region
but with different weights given to the local deviations of the halo
potential. It is possible that the different sampling has resulted in
the stars yielding a large scatter in the inferred halo properties.
To see whether this is the case, we select dark matter particles that
have the same $E$-$L$ distribution as the stars to create a star-like
dark matter sample.  Subhalo particles are removed from the DM and
star samples before sampling the $E$-$L$ distribution and the
radial coordinates are ignored when constructing the samples.  The
same fitting procedure is then applied to this star-like dark matter
sample. The results are shown in Fig.~\ref{fig:StarFitDM}. By drawing
a sample with the stellar $P(E,L)$ from DM particles, the scatter in
the fits is actually slightly decreased (probably due to the removal
of the less virialized outer halo particles) compared to the DM fits,
and is much smaller than that in the star fits. This shows that the
tagged stars are indeed in less of a steady state than the dark matter.

We remark that even though for the star samples the scatter in
Fig.~\ref{fig:TSProfStar} is higher at large radii, low binding
energies and high angular momentum, we do \emph{not} detect a
systematic decrease in the biases of the fitted parameters as we
exclude the large radii, low energy or low angular momentum regions.
The lack of systematic improvements in bias when excluding the regions
with large scatter is consistent with our previous argument that the
biases are limited by the effective number of independent streams. Also note the
bias and scatter are separate quantities, and we have observed that
in the high scatter regions the mean profile does not appear more biased.

\section{Half mass constraint}\label{sec:Mhalf}

\begin{figure}
 \myplot{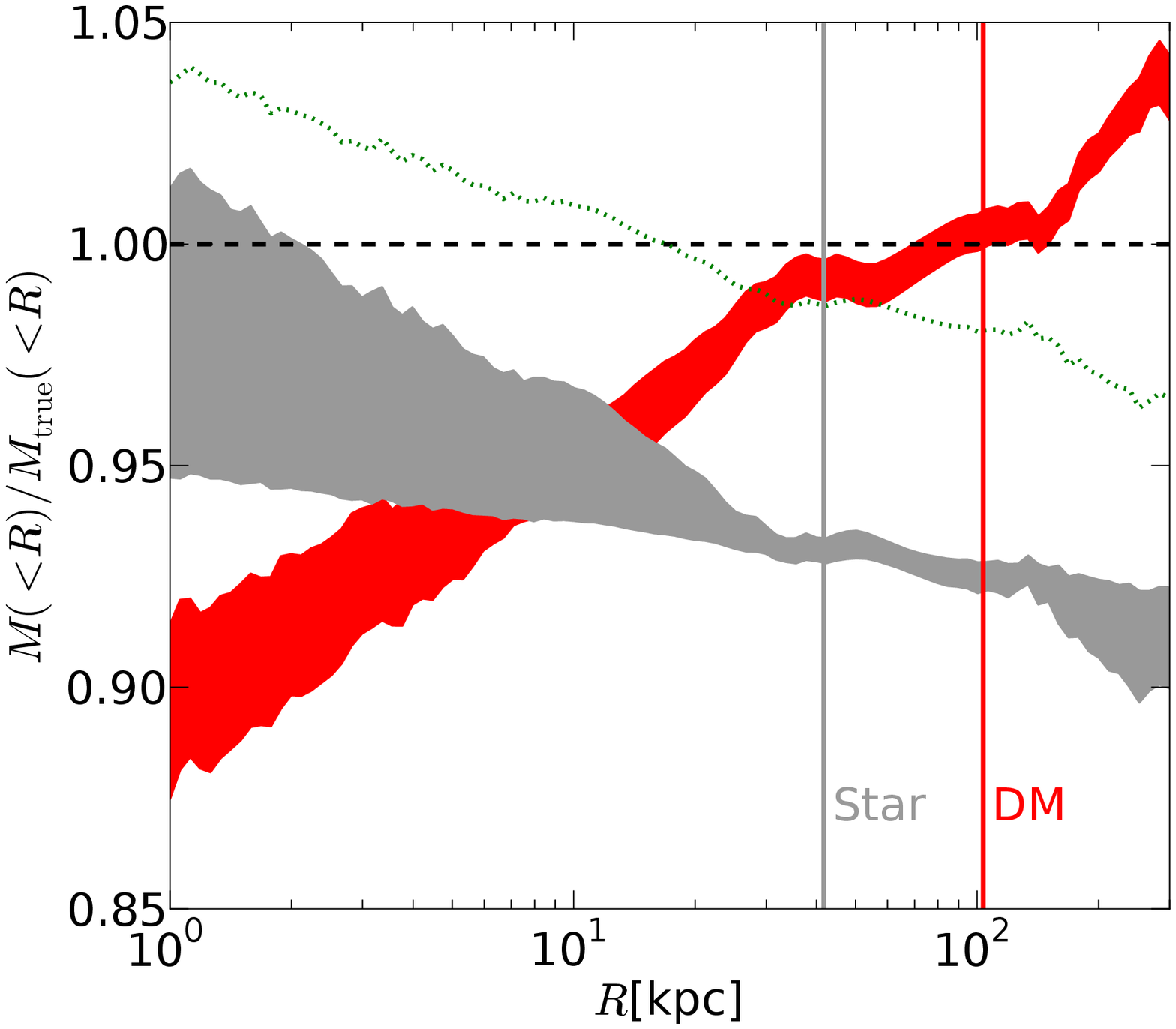}
 \caption{The template-fitted mass profile of halo~A, using smooth DM
 (red) and smooth star (grey) particles as tracers respectively. The shaded
 regions are the $1$-$\sigma$ constraints on the mass profile, normalized
 by the true profile. The vertical lines mark the half-mass radius of
 the two tracers. The green dotted line is the best-fitting profile
 from the star-like DM tracer (i.e., ``dmStar'' in
 Fig.~\ref{fig:StarFitDM}).}\label{fig:MassProf}
\end{figure}

In Paper~I we used Monte-Carlo samples to demonstrate that the mass
profiles are best constrained near the median radius of the tracer
population. \hl{Similar best-constrained masses also exist in several previous studies using very different methodologies~\citep[][see section 5.1 of Paper~I for more discussion]{Walker09,Wolf10,AE11}.} Here we revisit this discovery with the Aquarius
haloes. In Fig.~\ref{fig:MassProf} we plot the constrained mass
profile from the DM and star tracers in halo~A.  Note our likelihood
method constrains not only the characteristic mass, but also the shape
of the mass profile. If the profile shape is biased then the bias in
enclosed mass varies with radius. Consistent with our previous
findings, the mass is best
constrained near the half-mass radius of the tracer. Comparing these
best-constrained masses, the bias in stars is still significantly
larger than that in DM. This is also consistent with our test using a
star-like DM tracer in Fig.~\ref{fig:StarFitDM}, where we find that
the different samplings of the star and DM tracers are not the cause of
the different bias levels. For comparison, the best-fitting profile of
the star-like DM tracer is also plotted, and shows a bias comparable
to the original DM fit. As listed in Table~\ref{table:HalfBias}, stars
yield an average bias of $\sim 5\%$ at $r_{1/2}$ when using the
template fits, while the DM yields only $\sim 1\%$. The star-like DM
tracers have $r_{1/2}$ close to that of the stars, but gives almost no
bias at $r_{1/2}$. As far as the constraints at $r_{1/2}$ are
concerned, fitting with NFW profiles gives quite similar results to
template fits, indicating that the half mass constraint is less
sensitive to the adopted functional form for the halo density
profile (see also Paper~I). However, models with extra assumptions could still lead to
significant bias at $r_{1/2}$. For example, fitting the DM tracers
with the $f(E,L)$ method in \citet{Wang} produces an average mass bias
of $13\%$ at $r_{1/2}$ (Table~\ref{table:summary}). 

\begin{table*}
\caption{Tracer half-mass radius $r_{1/2}$ and mass bias $b_{1/2}$ at
$r_{1/2}$, from fits to the smooth component of DM, star, and
``dmStar`` tracers. ``dmStar'' refers to DM tracers selected to have
the same $E$-$L$ distribution as the stars, as in
Fig.~\ref{fig:StarFitDM}. We list the biases from template fits by
default. For DM and star tracers, biases from NFW fits are also given
in parenthesis.}\label{table:HalfBias}
\begin{center}
\begin{tabular}{ccccccc}
\hline
\hline Halo & DM $r_{1/2}/$kpc & DM $b_{1/2}$ & Star $r_{1/2}/$kpc & Star $b_{1/2}$ & dmStar $r_{1/2}$ & dmStar $b_{1/2}$\\
\hline  A2 & 103 & 0.00 (0.05) & 41.7 & -0.07 (-0.02) & 41.6 & -0.01\\
\hline  B2 & 85.4 & 0.01 (0.03) & 18.8 & 0.05 (0.05) & 19.0 & 0.01\\
\hline  C2 & 86.2 & 0.03 (0.05) & 48.2 & 0.03 (0.04)& 47.4 & 0.00\\
\hline  D2 & 103.1 & -0.02 (0.00) & 32.8 & 0.04 (0.05) & 33.1 & 0.00\\
\hline  E2 & 90.1 & -0.00 (0.00) & 18.6 & 0.04 (0.02) & 18.6 & 0.00\\
\hline
\end{tabular}
\end{center}
\end{table*}

We emphasize that since we are not only interested in the mass
constraint at a single radius but also in the full profile, any
parametrization of a specific density profile should be equivalent. As
long as the constraints are fully described in terms of the parameter
covariance or the 2-dimensional confidence contour, the constraints on
the full profile can always be recovered and translated to constraints
in any other parametrization of the profile.  Our parametrization is
intentionally chosen to constrain the most popular parameters, the
virial mass and concentration of haloes.

\section{Summary and conclusions}\label{sec:summary}

\begin{table*}
\caption{Summary of the different fits to the halo density profile. For each combination of data
and method, we list the fitted parameters averaged over the five
haloes ($\bar{x}$) and their halo-to-halo standard deviation
($\sigma$) in the form $\bar{x}\pm\sigma$. The mass ($M$) and
concentration ($c$) parameters are normalized by their true values,
$M_0$ and $c_0$. The mass bias at the tracer half-mass radius,
$b_{1/2}$, is also listed in the same form. Different columns refer to
different combinations of data and methods. ``DM-Full'', ``DM'' and
``Star'' refer to full DM, the smooth DM (DM-Full excluding subhalo
particles), and smooth star tracers. ``dmStar'' refers to DM samples
selected to have the same $E$-$L$ distribution as stars. ``NFW'' and
``TMP'' refer to fits using NFW or template potential
profiles. $f(E,L)$ refers to the $(r,v_r,v_t)$ fit in \citet{Wang}
using an $f(E,L)$ distribution function.}\label{table:summary}
\begin{center}
\begin{tabular}{ccccccccc}
\hline
\hline          & DM:NFW & DM:TMP & DM-Full:TMP &DM:$f(E,L)$  &Star:NFW & Star:TMP & Star:$f(E,L)$ &dmStar:TMP \\
\hline $M/M_0$  &$0.97\pm0.07$&$1.00\pm0.05$&$1.02\pm0.04$& $1.54\pm0.05$ & $1.00\pm0.25$ &$1.10\pm0.23$ & $0.82\pm0.16$&$0.99\pm0.03$\\
\hline $c/c_0$  &$0.99\pm0.17$&$0.98\pm0.06$&$0.96\pm0.04$& $0.48\pm0.13$ & $0.99\pm0.16$ &$0.97\pm0.18$ & $1.12\pm0.14$&$1.01\pm0.03$\\
\hline $b_{1/2}$&$0.02\pm0.02$&$0.00\pm0.01$ &$-0.01\pm0.02$ & $0.13\pm0.02$ & $0.03\pm0.02$& $0.02\pm0.04$&$-0.02\pm0.01$&$0.00\pm0.01$\\
\hline
\end{tabular}
\end{center}
\end{table*}

We have applied our oPDF estimator to tracers in five simulated haloes
from the Aquarius project, to study the level of systematic biases in
fitting the mass distribution of Milky-Way sized haloes. We focus on effects from the
parametrization of the halo potential, the existence of subhaloes, and
the types of tracer used. Assuming a spherical symmetric potential,
our method only makes use of the dynamical equilibrium of the tracer. As a
result, the level of systematic biases detected in our analysis can,
in general, be interpreted as the minimum level of bias present in any
time-independent DF modelling of dynamical tracers that assumes a
spherical potential. With our sample of five haloes, we do not have a
reliable detection of a common bias in our method towards any particular
direction in parameter space. Instead, we focus on characterizing
the average amplitude of the bias in each fit. We quantify this as the rms
scatter of the biases for individual haloes, and summarize them in
Table~\ref{table:summary}. The method works very well on DM tracers,
with a level of systematic bias at only $\sim 5\%$. Assuming an NFW
profile does not significantly affect the fits in most
cases, except in one case out of five where the density profile of the
halo (Aq-A) differs significantly from the NFW form, leading to a much
larger bias ($\sim 30\%$) when adopting the NFW profile.  However, the
deviation from the NFW profile in halo~A does \emph{not} affect the
equilibrium of the DM tracer. Subhaloes exist as perturbations that give rise to deviations from steady state for the tracers, but only affect the dynamical fits using DM tracers by $\sim 1\%$. In contrast to the fairly
good fits to the DM tracers with our method, a conventional DF fit
adopting \hl{a specific $f(E,L)$ DF~\citep{Wang}} yields a net bias of
$50\%$ in mass and concentration on average. 
This is caused by the additional assumptions made
in the $f(E,L)$ DF that restricts the allowed distribution of
orbits. On the other hand, the halo-to-halo variation in the bias is
comparable to that in our method, demonstrating that our method gives 
the minimum uncertainty in mass modelling assuming time-independent
DFs.

Applying our method to mock stars results in a higher level of bias,
$b\sim 20\%$, comparable to that in the $f(E,L)$ DF method tested in \citet{Wang} which, 
however, also suffers from a non-zero net bias. 
The larger bias using star tracers is not due to
different phase-space sampling by stars compared to DM particles: DM
tracers constrained to sample the phase space $E$-$L$ distribution
in the same way as the stars yield biases at the same level as the original DM tracers. The
larger deviation of tagged stars from a steady state is not surprising
because they involve the 1\% most-bound particles of their
host subhalo at the time of star formation. By definition, these
particles are the most resistant to tidal stripping and subsequent mixing. Even though we
only use the stripped population of tagged particles, they are
still farther from equilibrium than the smooth component of
the host halo.

It is well known that dynamical tracers best-constrain the host mass
near the tracer's half-mass radius,
$r_{1/2}$~\citep{Walker09,Wolf10,AE11}. Although we adopt a vastly different method from those analysis, a similar behaviour is also observed in
our analysis. Near $r_{1/2}$, the mass biases, $b_{1/2}$, are much
reduced, and also become less sensitive to the functional form of the halo
profile assumed in our model. Larger biases are still observed for
stars, $b_{1/2} \sim 5\%$, compared with $b_{1/2} \sim 2\%$ for
DM tracers. The $f(E,L)$ method that poorly fits the DM tracer
produces a much larger bias, $b_{1/2} \sim 10\%$. Although the bias at $r_{1/2}$
is significantly smaller than the bias for the total mass, in reality $b_{1/2}$ together with the
constraint on the profile shape at $r_{1/2}$ is equivalent to the joint constraint in the mass-concentration space. Given the full
mass-concentration covariance matrix, one can readily obtain the mass
constraint at any radius including $r_{1/2}$. While $r_{1/2}$ depends on the tracer, $M$ and $c$ are intrinsic properties of the halo.

\hl{In this work we frequently compare our results with those of \citet{Wang} who used a specific model of the $f(E,L)$ family to study the same haloes. We demonstrate that the extra assumptions in that model beyond time-independence and spherical symmetry have resulted in a worse performance compared with the oPDF. There are more flexible distribution functions that can improve over the one assumed in \citet{Wang}, for example, by allowing for varying anisotropies~\citep[e.g.][]{Wojtak,WE15}. More generally, there may exist a true model that describes well the distribution function of the tracers. However, such a true model has to be known \textit{a priori} to fit the tracers correctly, which is a highly challenging task if at all possible. At the same time, any specifically proposed distribution function has generally has limitations stemming from extra assumptions over and above the Jeans theorem. These extra assumptions may not be obeyed by an arbitrary tracer sample. As such, the results obtained using the oPDF method which makes minimal assumptions are particularly robust, and the comparison with the specific model of \citet{Wang} therefore serves to illustrate the limitations of restricted models. In particular, since the statistical noise has been controlled to be negligible in our analysis, the level of systematic bias detected with the oPDF is the minimum level of systematic bias expected from any model that assumes a spherically symmetric and time-independent distribution function.}

Note that the stars used in this work are generated from a particle
tagging method that is a relatively simple way of approximating the
phase-space distribution of stars. Several factors, including
insufficient mass resolution (the weighting and multiple tagging of
star particles), the time discreteness of the tagging (the method
works with snapshots), and the lack of dissipation and back-reaction
on the potential from stars, could all potentially affect the degree of realism with which the tagged stars represent the dynamics of real stars. Due to these limitations, the results from the tagged stars should only be taken as indicative.
Also note that only five haloes are studied in this work and these may
not be very representative of our Milky Way halo. In a follow up work, we will apply the
method to a larger sample of haloes modelled using SPH simulations of the Local
Group for a more realistic assessment of the dynamical state of
tracers in Galactic haloes.

\section*{Acknowledgements}
We thank Julio Navarro, Andrew Pontzen, Yanchuan Cai and Vince Eke for helpful comments and discussions. 
This work was supported by the European Research Council [GA 267291] COSMIWAY and Science and Technology Facilities Council
Durham Consolidated Grant. WW acknowledges a Durham Junior Research Fellowship.
This work used the DiRAC Data Centric system at Durham University,
operated by the Institute for Computational Cosmology on behalf of the
STFC DiRAC HPC Facility (www.dirac.ac.uk). This equipment was funded
by BIS National E-infrastructure capital grant ST/K00042X/1, STFC
capital grant ST/H008519/1, and STFC DiRAC Operations grant
ST/K003267/1 and Durham University. DiRAC is part of the National
E-Infrastructure. This work was supported by the Science and
Technology Facilities Council [ST/F001166/1]. 

The code implementing our method is freely available at a GitHub repository linked from  \url{http://icc.dur.ac.uk/data/#oPDF}.

\bibliographystyle{\mybibstyle}
\setlength{\bibhang}{2.0em}
\setlength\labelwidth{0.0em}
\bibliography{fitting}

\appendix
\section{Parameters of template profiles}\label{app:template}
To connect template parameters $A$ and $B$ in Eq.~\eqref{eq:template} to physical parameters, we can define 
\begin{align}
 A&=\psi_{\rm s}/\psi_{\rm s0},\\
 B&=r_{\rm s}/r_{\rm s0},
\end{align} 
where $r_{\rm s}$ is a scale radius at which the profile has some
predefined shape, and $\psi_{\rm s}$ is the potential at $r=0$. We
choose $r_{\rm s}$ to be the radius where $d\ln \rho/d\ln r=-2$, to be
consistent with an NFW parametrization. $r_{\rm s0}$ and $\psi_{\rm
s0}$ are the corresponding quantities of the true profile. Hence this
template profile is parametrized by $(A,B)$ or equivalently by
$(\psi_{\rm s},r_{\rm s})$. We can also define equivalent mass and
concentration parameters.  For each profile, the virial mass, $M$, and
virial radius, $R_{\rm v}$, can be defined following the same
spherical-overdensity definition as in Eq.~\eqref{eq:vir}, and the
concentration can be defined through $c=R_{\rm v}/r_{\rm s}$,
consistently with NFW.\footnote{Although we have chosen $r_{\rm s}$ to
be the slope $-2$ radius, in principle $r_{\rm s}$ can be defined to
be the radius at any characteristic slope, with the concentration
parameter being interpreted as the ratio between $R_{\rm v}$ and
$r_{\rm s}$. As long as the definition is consistent within the same
template, the $B$ parameter does not depend on the specific definition
of $r_{\rm s}$.} The mass and concentration parameters of the true
profile (i.e., the template with $A=1,B=1$), $M_0$ and $c_0$, are by
definition the \emph{true parameters} of the halo, and can be obtained
unambiguously from the true profile without fitting. If the halo is
perfectly NFW, then the true parameters defined this way are also the
best-fitting NFW parameters to the density profile. When the density
profile differs from NFW form, however, the true parameters, $M_0$ and
$c_0$, should be interpreted as the spherical overdensity mass and the
contrast of the spherical overdensity radius, $R_{\rm v}$, to the
slope $-2$ radius, $r_{\rm s}$, rather than being any best-fitting NFW
parameters.

With the template parametrization, the inversion from any set of
($M,c$) parameters back to ($A,B$) is also straight-forward. Note that
the mass profile of the template scales as
\begin{align}
M(r)&=\frac{\psi'(r)r^2}{G}\\\nonumber
&=ABm \left(\frac{r}{B}\right),
\end{align} 
where $M(r)$ is the mass profile of the template with parameters
($A,B$), $\psi'(r)$ is the derivative of the template potential, and
$m(r)$ is the true mass profile. Hence $M=ABm(R_{\rm v}/B)$. After
obtaining ($R_{\rm v},r_{\rm s}$) from ($M,c$), one can solve ($A,B$)
as follows
\begin{align}
 B&=\frac{r_{\rm s}}{r_{\rm s0}}\\
 A&=\frac{M}{Bm(R_{\rm v}/B)}.\\
\end{align}
To create the templates numerically, we extract both the potential
profiles and the cumulative density profiles $\rho(<r)\propto
{\psi'(r)}/{r}$ from the particle distribution of each halo. The
$\rho(<r)$ is provided to avoid the need for numerical differentiation
of the potential profile. 

%

\end{document}